\documentclass[12pt]{article}
\usepackage[T1]{fontenc}
\usepackage[utf8]{inputenc}
\usepackage{authblk}

\title{\LARGE Effect estimation in the presence of a misclassified binary mediator}

\author[]{Kimberly A. Hochstedler Webb}
\author[]{Martin T. Wells}
\affil[]{Department of Statistics and Data Science, Cornell University \\ Ithaca, NY}
\affil[]{kah343@cornell.edu}

\usepackage[super,sort&compress]{natbib}
\setcitestyle{comma,numbers,super,open={},close={}} 
\usepackage{graphicx}
\usepackage[paperwidth=8.5in,paperheight=11.0in,top=1in, bottom=1in, left=1in, right=1in,lines=25]{geometry}
\usepackage[title,toc,titletoc,page]{appendix}

\usepackage{siunitx}
\usepackage{appendix}
\usepackage{mathtools}
\usepackage{threeparttable}
\usepackage{booktabs}
\usepackage{bbm}
\usepackage{caption}
\usepackage{algpseudocode}
\usepackage{algorithm}
\usepackage{url}
\usepackage{xcolor}
\usepackage{lscape}
\usepackage{subfiles}

\usepackage{amsmath}
\usepackage{blkarray}

\usepackage{setspace}
\DeclareMathOperator{\logit}{logit}
\usepackage{stix}

\begin{document}

\maketitle

\begin{abstract}
Mediation analyses allow researchers to quantify the effect of an exposure variable on an outcome variable through a mediator variable. If a binary mediator variable is misclassified, the resulting analysis can be severely biased. Misclassification is especially difficult to deal with when it is differential and when there are no gold standard labels available. Previous work has addressed this problem using a sensitivity analysis framework or by assuming that misclassification rates are known. We leverage a variable related to the misclassification mechanism to recover unbiased parameter estimates without using gold standard labels. The proposed methods require the reasonable assumption that the sum of the sensitivity and specificity is greater than 1. Three correction methods are presented: (1) an ordinary least squares correction for Normal outcome models, (2) a multi-step predictive value weighting method, and (3) a seamless expectation-maximization algorithm. We apply our misclassification correction strategies to investigate the mediating role of gestational hypertension on the association between maternal age and pre-term birth.

\textbf{Keywords: } mediation analysis, causal effects, bias-correction, label switching, predictive value weighting, EM algorithm

\end{abstract}

\newpage


\section{Introduction}
Mediation analysis quantifies the effect of an exposure on an outcome, through an intermediate or mediator variable. A mediation model allows for the exposure to have a direct association with the outcome of interest in addition to an indirect association. For the indirect association, the mediation model hypothesizes that the exposure may impact the mediator variable, which in turn is associated with the outcome.\cite{baron1986moderator} Mediation analyses are popular in social science and public health association studies, and have more recently been utilized to study causal relationships.\cite{jung2021introduction} In particular, the use of the potential outcomes framework has allowed for the study of direct and indirect effects of the exposure on the outcome, through the mediator.\cite{imai2011commentary}

An implicit assumption in many mediation methods is that the variables used in the analysis are measured without error. However, if this assumption is violated through measurement error in the exposure, mediator, and/or outcome variables, the resulting causal effect estimates may be biased.\cite{li2020direct} Jiang and VanderWeele (2015) find that non-differential misclassification of a binary outcome variable results in effect estimates that are biased toward the null.\cite{jiang2015causal} Non-differential exposure misclassification, on the other hand, may lead to over- or under-estimation of causal effects.\cite{valeri2017misclassified} Similarly, non-differential misclassification of a binary mediator has been shown to lead to overestimation of the natural direct effect and underestimation of the natural indirect effect.\cite{ogburn2012analytic} Measurement error in continuous mediator variables also results in bias in the resulting effect estimates.\cite{le2012quantification}

Other works have studied the impact of differential misclassification in mediation analysis. When an outcome variable is differentially misclassified, as when the misclassification rates depend on the exposure variable, the resulting bias in the odds ratio and risk ratio depends on some sensitivity parameters in the system.\cite{vanderweele2019simple} An extension of this result showed that differential misclassification in the exposure, mediator, and/or outcome variables can yield estimated direct effects that are biased non-intuitive ways.\cite{li2020direct}

In a regression-based mediation analysis, parameters from regressions of the outcome variable and mediator variable are used to compute causal effects.\cite{li2007estimation, valeri2013mediation} Thus, the bias in causal effects documented in previous work can be understood by investigating the bias in regression parameter estimates when a binary mediator variable is misclassified.\cite{valeri2014estimation} Valeri and VanderWeele (2014) demonstrate how the asymptotic bias of parameter estimates from mediator and outcome regression models fit with misclassified binary mediators induce bias in causal effect estimates.\cite{valeri2014estimation} Moreover, they propose bias-correction strategies for cases with a non-differentially misclassified binary mediator. These correction strategies rely on known mediator sensitivity and specificity values, or on the use of a sensitivity analysis approach.\cite{valeri2014estimation}

In more general regression frameworks, bias correction strategies have been proposed for differentially misclassified binary exposure and/or outcome variables. Numerous likelihood-based approaches rely on the availability of a gold-standard instrument for validation samples.\cite{lotspeich2022efficient, tang2015binary, tao2021efficient} Other correction methods rely on known misclassification rates or suggest a sensitivity analysis approach.\cite{nguimkeu2021regression, lyles2010sensitivity, magder1997logistic, valeri2014estimation} For example, Lyles and Lin (2010) suggest a predictive value weighting approach for misclassified binary covariates when investigator-supplied sensitivity and specificity values are available.\cite{lyles2010sensitivity} Nguimkeu, Roseman, and Tennekoon (2020) derive asymptotic bias expressions for ordinary least squares parameter estimates obtained with a misclassified binary regressor. They use these expressions to define bias-adjusted parameter estimates using given misclassification probabilities.\cite{nguimkeu2021regression}

When a gold standard instrument or investigator-supplied misclassification rates are unavailable, Beesley and Mukherjee (2020) develop methods for inference in the presence of binary outcome misclassification. These methods leverage patient-related factors associated with the misclassification mechanism to correct association parameter estimates, but require an assumption of perfect specificity in the outcome.\cite{beesley2020statistical} Webb and Wells (2023) extend these methods to allow for both imperfect sensitivity and imperfect specificity in the outcome variable.\cite{hochstedler2023statistical} Webb, Riley, and Wells (2023) further extend these bias-correction procedures to account for multiple, sequential misclassified binary outcomes.\cite{webb2023assessment}

In this paper, we develop new strategies for handling misclassification in mediator variables, in the absence of a gold standard instrument or investigator-supplied sensitivity and specificity values. We propose three novel bias-correction procedures to estimate both the rate of misclassification in the mediator and the unbiased association parameters in the mediator and outcome regression models. These methods incorporate covariates related to the misclassification mechanism in order to minimize external information required for the analysis. All of the proposed methods are available in an open-source R Package called \textit{COMMA} (COrrecting Misclassified Mediation Analysis). \textit{COMMA} can also fit models with Normal, Poisson, or Binary outcome distributions.~\cite{COMMA} Through simulation studies, we demonstrate the utility of these methods in comparison to procedures that ignore misclassification in the binary mediator variable. We apply our proposed methods to study the potential mediating relationship of gestational hypertension on the association between maternal age and pre-term delivery. 

This paper is organized as follows. Section \ref{model} outlines the proposed model and conceptual framework. In Section \ref{estimation-methods}, we propose three novel methods to account for a misclassified binary mediator variable. These methods include a bias-corrected least squares estimator, a predictive value weighting procedure, and a seamless EM algorithm. In Section \ref{simulations}, we evaluate the performance of the proposed methods via simulation study. In Section \ref{example}, we apply two of the proposed methods to a study of risk factors for pre-term delivery. We conclude with a discussion in Section \ref{discussion}.

\section{Model, Notation, and Conceptual Framework} \label{model}
Let $X$ be a predictor of interest, $Y$ be an outcome, $M$ be a binary mediator, and $\boldsymbol{C}$ be a matrix of continuous or categorical covariates. Suppose that $M$ is a latent or unobserved variable that can take values $j \in \{1,2\}$. Let $M^*$ denote a potentially misclassified, observed version of the mediator variable, $M$. $M^*$ can take values $\ell \in \{1,2\}$. Let $\boldsymbol{Z}$ denote a matrix of covariates related to the sensitivity and specificity of mediator variable. We assume that $X$, $Y$, $\boldsymbol{C}$, and $\boldsymbol{Z}$ are measured without error. We use the following models to relate the variables in this system: 
\begin{align}
\label{eq:true-mediator}
\text{True mediator mechanism: } &\; \logit\{ P(M = 1 | X, \boldsymbol{C} ; \boldsymbol{\beta}) \} = \beta_{0} + \beta_{X} X + \boldsymbol{\beta_{C} C}.\\
\label{eq:observed-mediator-1}
\text{Observed mediator mechanisms: } &\; \logit\{ P(M^* = 1 | M = 1, \boldsymbol{Z} ; \boldsymbol{\gamma}) \} = \gamma_{110} + \boldsymbol{\gamma_{11Z} Z}, \\
\label{eq:observed-mediator-2}
                                &\; \logit\{ P(M^* = 1 | M = 2, \boldsymbol{Z} ; \boldsymbol{\gamma}) \} = \gamma_{120} + \boldsymbol{\gamma_{12Z} Z}. \\
\label{eq:outcome-mechanism}
\text{Outcome mechanism: } &\;E(Y | X, \boldsymbol{C}, M; \boldsymbol{\theta}) = \theta_0 + \theta_X X + \boldsymbol{\theta_C C} + \theta_M M + \theta_{XM} XM.
\end{align}
If we have a Bernoulli outcome that we model with a logit link, the \textit{outcome mechanism} in (\ref{eq:outcome-mechanism}) can be replaced by 
\begin{equation}
\begin{aligned}
\label{eq:bernoulli-outcome-mechanism}
\logit \{P(Y = 1 | X, \boldsymbol{C}, M; \boldsymbol{\theta}) \} = \theta_0 + \theta_X X + \boldsymbol{\theta_C C} + \theta_m M + \theta_{XM} XM.
\end{aligned}
\end{equation}
If we instead have a Poisson outcome that we model with a log link, the \textit{outcome mechanism} in (\ref{eq:outcome-mechanism}) can be replaced by 
\begin{equation}
\begin{aligned}
\label{eq:poisson-outcome-mechanism}
\log \{E(Y | X, \boldsymbol{C}, M; \boldsymbol{\theta}) \} = \theta_0 + \theta_X X + \boldsymbol{\theta_C C} + \theta_m M + \theta_{XM} XM.
\end{aligned}
\end{equation}
The model in (\ref{eq:true-mediator}) - (\ref{eq:outcome-mechanism}) allows for differential and covariate-dependent misclassification of the mediator. That is, the probability of observing a given value of the mediator, $M^*$ depends on both the value of the true mediator and on the values of predictors in $\boldsymbol{Z}$. In this model, the \textit{true mediator mechanism} relates the predictor of interest $X$ and the covariate matrix $\boldsymbol{C}$ to the unobserved binary mediator $M$ using a vector of regression parameters denoted $\boldsymbol{\beta}$. The \textit{observed mediator mechanism} models the association between $\boldsymbol{Z}$ and the observed mediator, $M^*$, conditional on the true unobserved mediator $M$, as a function of the vector of parameters $\boldsymbol{\gamma}$. Note that the component of the \textit{observed mediator mechanism} with $M = 1$ governs the sensitivity of the mediator measure, while the component with $M = 2$ governs the specificity of the mediator measure. Finally, the \textit{outcome mechanism} relates the predictor of interest $X$, the covariate matrix $\boldsymbol{C}$, and the true unobserved mediator $M$ to the outcome of interest, $Y$, using the vector of regression parameters $\boldsymbol{\theta}$. In the \textit{true mediator mechanism} and the \textit{observed mediator mechanism}, $M = 2$ and $M^* = 2$ are the respective reference categories, and all corresponding $\boldsymbol{\beta}$ and $\boldsymbol{\gamma}$ parameters are set to 0. 

We can rearrange the \textit{true mediator mechanism} and \textit{observed mediator mechanism} in (\ref{eq:true-mediator}) - (\ref{eq:observed-mediator-2}) to express response probabilities for subject $i$'s true mediator value and for subject $i$'s observed mediator category, given the true mediator, as follows:
\begin{align}
\label{eq:pi-m}
P(M_i = 1 | X_i = x, \boldsymbol{C_i} ; \boldsymbol{\beta}) = &\; \; \pi_{i1} = \frac{\text{exp}\{\beta_{0} + \beta_{X} x_i + \boldsymbol{\beta_{C} C_i} \} }{1 + \text{exp}\{\beta_{0} + \beta_{X} x_i + \boldsymbol{\beta_{C} C_i}\}} \\
\label{pistar-m}
P(M^*_i = 1 | M_i = j, \boldsymbol{Z} ; \boldsymbol{\gamma}) = &\; \pi^*_{i1j} = \frac{\text{exp}\{\gamma_{1j0} + \boldsymbol{\gamma_{1jZ} Z_i}\}}{1 + \text{exp}\{\gamma_{1j0} + \boldsymbol{\gamma_{1jZ} Z_i}\}}.
\end{align}
We can compute these quantities for all subjects $i = 1, \dots, N$ in the sample. When $j = 1$ and $\ell = 1$, the quantity $\pi^*_{11} = \frac{1}{N}\sum_{i = 1}^{N} \pi^*_{i11}$ expresses the average sensitivity for the mediator measure. When $j = 2$ and $\ell = 2$, $\pi^*_{22} = \frac{1}{N}\sum_{i = 1}^{N} \pi^*_{i22}$ denotes the average specificity. 

If we were to use the misclassified mediator $M^*$ in place of the true mediator, $M$, in a standard regression framework for mediation, we would obtain the following \textit{analysis model}:
\begin{align}
\label{eq:analysis-model}
&\logit\{ P(M^* = 1 | X, \boldsymbol{C} ; \boldsymbol{\beta}^*) \} = \beta^*_{0} + \beta^*_{X} X + \boldsymbol{\beta^*_{C} C} 
\\
&E(Y | X, \boldsymbol{C}, M^*; \boldsymbol{\theta}^*) = \theta^*_0 + \theta^*_X X + \boldsymbol{\theta^*_C C} + \theta^*_M M^* + \theta_{XM^*} XM^*.
\end{align}
Misclassification in $M^*$ generally results in bias of the parameter vectors $\boldsymbol{\beta^*}$ and $\boldsymbol{\theta^*}$ such that $\boldsymbol{\beta^*} \neq \boldsymbol{\beta}$ and $\boldsymbol{\theta^*} \neq \boldsymbol{\theta}$.\cite{valeri2014estimation} Our objective is to estimate $\boldsymbol{\beta}$ and $\boldsymbol{\theta}$, the true mediator and outcome regression parameters in (\ref{eq:true-mediator}) and (\ref{eq:outcome-mechanism}). We can view the \textit{observation mechanisms} in (\ref{eq:observed-mediator-1}) and (\ref{eq:observed-mediator-2}) as a secondary interest, describing the relationship between $\boldsymbol{Z}$ and misclassification of the mediator $M$.

\subsection{Estimation of Causal Effects}
For parameters $\boldsymbol{\beta}$ and $\boldsymbol{\theta}$ in (\ref{eq:true-mediator}) and (\ref{eq:outcome-mechanism}), we can write the natural direct effect NDE$(\boldsymbol{\beta}, \boldsymbol{\theta})$, natural indirect effect NIE$(\boldsymbol{\beta}, \boldsymbol{\theta})$, and controlled direct effect CDE$(\boldsymbol{\beta}, \boldsymbol{\theta})$ for a change in predictor $X$ from $\Tilde{x}$ to $x$ as\cite{valeri2013mediation}
\begin{align}
\label{eq:continuous-effect-est}
& CDE=\left(\theta_X+\theta_{XM} m\right)\left(x - \Tilde{x} \right) \\
& NDE=\theta_X\left(x-\Tilde{x}\right)+\left\{\theta_{XM}\left(x - \Tilde{x} \right)\right\} \frac{\exp \left(\beta_0+\beta_X \Tilde{x}+\beta_C c\right)}{1+\exp \left(\beta_0+\beta_X \Tilde{x}+\beta_C c\right)} \\
& NIE=\left(\theta_M+\theta_{XM} x\right)\left\{\frac{\exp \left(\beta_0+\beta_X x+\beta_C c\right)}{1+\exp \left(\beta_0+\beta_X x + \beta_C c\right)}-\frac{\exp \left(\beta_0+\beta_X \Tilde{x} +\beta_C c\right)}{1+\exp \left(\beta_0+\beta_X \Tilde{x}+\beta_C c\right)}\right\}.
\end{align}

If the outcome is binary and rare, we use $\boldsymbol{\beta}$ and $\boldsymbol{\theta}$ from (\ref{eq:true-mediator}) and (\ref{eq:bernoulli-outcome-mechanism}) to compute NDE$(\boldsymbol{\beta}, \boldsymbol{\theta})$, natural indirect effect NIE$(\boldsymbol{\beta}, \boldsymbol{\theta})$, and controlled direct effect CDE$(\boldsymbol{\beta}, \boldsymbol{\theta})$ for a change in predictor $X$ from $\Tilde{x}$ to $x$ in terms of odds ratios, \cite{valeri2013mediation}
\begin{align}
\label{eq:rare-binary-effect-est}
& OR^{CDE}=\exp \left\{\left(\theta_X+\theta_{XM} m\right)\left(x-\Tilde{x}\right)\right\} \\
& OR^{NDE} \cong \frac{\exp \left(\theta_X x\right)\left\{1+\exp \left(\theta_M+\theta_{XM} x+\beta_0+\beta_X \Tilde{x}+\beta_C c\right)\right\}}{\exp \left(\theta_X \Tilde{x}\right)\left\{1+\exp \left(\theta_M+\theta_{XM} \Tilde{x} +\beta_0+\beta_X \Tilde{x} +\beta_C c\right)\right\}} \\
& OR^{NIE} \cong \frac{\left.\left\{1+\exp \left(\beta_0+\beta_X \Tilde{x}+\beta_Cc\right)\right\} 1+\exp \left(\theta_X+\theta_{XM} x+\beta_0+\beta_X x+\beta_C c\right)\right\}}{\left\{1+\exp \left(\beta_0+\beta_X x+\beta_C c\right)\right\}\left\{1+\exp \left(\theta_M+\theta_{XM} x+\beta_0+\beta_X \Tilde{x} +\beta_C c\right)\right\}}.
\end{align}

For count outcomes with a log link in the \textit{outcome mechanism}, as in (\ref{eq:poisson-outcome-mechanism}), effects are computed on the risk ratio scale using  $\boldsymbol{\beta}$ and $\boldsymbol{\theta}$ as follows:
\begin{align}
\label{eq:poisson-effect-est}
& RR^{CDE}=\exp \left\{\left(\theta_X+\theta_{XM} m\right)\left(x-\Tilde{x}\right)\right\} \\
& RR^{NDE} \cong \frac{\exp \left(\theta_X x\right)\left\{1+\exp \left(\theta_M+\theta_{XM} x+\beta_0+\beta_X \Tilde{x}+\beta_C c\right)\right\}}{\exp \left(\theta_X \Tilde{x}\right)\left\{1+\exp \left(\theta_M+\theta_{XM} \Tilde{x} +\beta_0+\beta_X \Tilde{x} +\beta_C c\right)\right\}} \\
& RR^{NIE} \cong \frac{\left.\left\{1+\exp \left(\beta_0+\beta_X \Tilde{x}+\beta_Cc\right)\right\} 1+\exp \left(\theta_X+\theta_{XM} x+\beta_0+\beta_X x+\beta_C c\right)\right\}}{\left\{1+\exp \left(\beta_0+\beta_X x+\beta_C c\right)\right\}\left\{1+\exp \left(\theta_M+\theta_{XM} x+\beta_0+\beta_X \Tilde{x} +\beta_C c\right)\right\}}.
\end{align}

\section{Estimation Methods} \label{estimation-methods}
In this section, we describe three approaches for estimating our proposed model in (\ref{eq:true-mediator}) - (\ref{eq:outcome-mechanism}). First, we describe a two-step ordinary least squares (OLS) correction procedure suitable for cases with a continuous outcome $Y$, based on the work of Nguimkeu, Rosenman, and Tennekoon (2021).\cite{nguimkeu2021regression} Next, we present a two-step extension of the predictive value weighting method first described in Lyles and Lin (2010).\cite{lyles2010sensitivity} Finally, we propose a seamless expectation-maximization (EM) algorithm approach for correcting for mediator misclassification and estimating the outcome model in a single step. All proposed estimation methods are available in the R package \textit{COMMA} for Normal, Poisson, and Binary outcome models. \cite{COMMA}

\subsection{Ordinary Least Squares Correction} \label{ols-correction}
We estimate the \textit{true mediator mechanism} and the \textit{observed mediator mechanisms} using the EM algorithm approach described in Webb and Wells (2023).\cite{hochstedler2023statistical} Considering only mechanisms (\ref{eq:true-mediator}) - (\ref{eq:observed-mediator-2}) and viewing the true mediator $M$ as a latent variable, we begin by constructing the complete data log-likelihood as 
\begin{equation}
    \begin{aligned}
    \label{eq:complete-log-like-combo}
    \ell_{complete}(\boldsymbol{\beta},& \boldsymbol{\gamma}; X, \boldsymbol{C}, \boldsymbol{Z}) \\
    &= \sum_{i = 1}^N \bigg[ \sum_{j = 1}^2 m_{ij} \text{log} \{ P(M_i = j | X_i, \boldsymbol{C_i}) \} + \sum_{j = 1}^2 \sum_{\ell = 1}^2 m_{ij} m^*_{i\ell} \text{log} \{ P(M^*_i = \ell | M_i = j, \boldsymbol{Z_i}) \}\bigg] & \\
    &= \sum_{i = 1}^N \bigg[ \sum_{j = 1}^2 m_{ij} \text{log} \{ \pi_{ij} \} + \sum_{j = 1}^2 \sum_{\ell = 1}^2  m_{ij} m^*_{i\ell} \text{log} \{ \pi^*_{i \ell j} \}\bigg],
    \end{aligned}
\raisetag{12pt}\end{equation}
where $m_{ij} = \mathbbm{I}(M_i = j)$ and $m^*_{i\ell} = \mathbbm{I}(M^*_i = \ell)$.

Since (\ref{eq:complete-log-like-combo}) is linear in the latent variable, $m_{ij}$, we can replace this value with it's expectation in the E-step of the EM algorithm. The expectation of $m_{ij}$ is defined as
\begin{equation}
\begin{aligned}
\label{eq:e-step-combo}
w^W_{ij} = P(M_i = j | M_i^*, X_i, \boldsymbol{C}_i, \boldsymbol{Z}_i) = \sum_{\ell = 1}^2 \frac{m^*_{i\ell} \pi^*_{i\ell j} \pi_{ij}}{\sum_{k = 1}^2 \pi^*_{i \ell k} \pi_{ik}}.  
\end{aligned}
\end{equation}

For the M-step, we maximize the expectation of the complete data log-likelihood, 
\begin{equation}
\begin{aligned}
\label{eq:m-step}
Q^W = \sum_{i = 1}^N \Bigl[ \sum_{j = 1}^2 w^W_{ij} \text{log} \{ \pi_{ij} \} + \sum_{j = 1}^2 \sum_{\ell = 1}^2 w^W_{ij} m^*_{i\ell} \text{log} \{ \pi^*_{i\ell j} \}\Bigr],
\end{aligned}
\end{equation}
with respect to $\boldsymbol{\beta}$ and $\boldsymbol{\gamma}$. We follow the label switching correction procedure described in Webb and Wells (2023) to obtain our final parameter estimates for the \textit{true mediator mechanism} and the \textit{observed mediator mechanisms}.\cite{hochstedler2023statistical} We denote these estimates as $\hat{\boldsymbol{\beta}}^W$ and $\hat{\boldsymbol{\gamma}}^W$.

In order to estimate (\ref{eq:outcome-mechanism}), we follow the bias correction method presented in Nguimkeu, Rosenman, and Tennekoon (2021) for misclassified predictor binary variables and continuous outcomes.\cite{nguimkeu2021regression} These methods require fixed misclassification probabilities, which are assumed to be known or given. We extend the work of Nguimkeu, Rosenman, and Tennekoon (2021) by using the $\hat{\boldsymbol{\gamma}}^W$ coefficients to obtain estimates of misclassification probabilities, rather than using a known value. We compute the average mediator sensitivity and specificity across all subjects, denoted $\hat{\pi}^*_{11}$ and $\hat{\pi}^*_{22}$ respectively, in the dataset as follows:
\begin{align}
\label{eq:avg_pistar}
\hat{\pi}^{*W}_{11} = \frac{1}{N}\sum_{i = 1}^N \hat{\pi}^{*W}_{i11} = \frac{1}{N}\sum_{i = 1}^N \frac{\text{exp}\{\hat{\gamma}^W_{110} + \boldsymbol{\hat{\gamma}^W_{11Z} Z_i}\}}{1 + \text{exp}\{\hat{\gamma}^W_{110} + \boldsymbol{\hat{\gamma}^W_{11Z} Z_i}\}},\\
\hat{\pi}^{*W}_{22} = \frac{1}{N}\sum_{i = 1}^N \hat{\pi}^{*W}_{i22} = \frac{1}{N}\sum_{i = 1}^N \frac{1}{1 + \text{exp}\{\hat{\gamma}^W_{120} + \boldsymbol{\hat{\gamma}^W_{12Z} Z_i}\}}.
\end{align}

Denote $S_{AB} = \frac{1}{N} \sum_{i = 1}^N (A - \bar{A}) (B - \bar{B})^T$ as the sample covariance between variables $A$ and $B$. $\bar{A}$ and $\bar{B}$ denote the sample means of $A$ and $B$, respectively. Let $\boldsymbol{D}$ denote the matrix including both $X$ and $\boldsymbol{C}$ predictors in the \textit{outcome mechanism}. Importantly, the \textit{outcome mechanism} cannot include an interaction term between $X$ and $M$ when using the OLS correction method. The inability to include an interaction term between the predictor of interest and the potentially misclassified binary variable is a limitation of the existing method.\cite{nguimkeu2021regression}

The bias-corrected $\boldsymbol{\theta}$ estimates for all slope terms are computed as follows:
\begin{equation}
\label{ols-correction-system}
\begin{bmatrix} \hat{\theta}_M  \\ \hat{\theta}_{D} \end{bmatrix} = \begin{bmatrix} (1 - \zeta) S_{M^*M^*} & S_{M^*D}  \\ (1 + \xi) S_{DM^*} & S_{DD}  \end{bmatrix}^{-1}  \begin{bmatrix} S_{YM^*}  \\ S_{YD} \end{bmatrix}.
\end{equation}

The quantities $\zeta$ and $\xi$ are defined as,
\begin{align}
     &\zeta = 1 - \frac{(\pi^*_1 - \hat{\pi}^{*W}_{12}) (1 - \hat{\pi}^{*W}_{21} - \pi^*_{1})}{(1 - \hat{\pi}^{*W}_{12} - \hat{\pi}^{*W}_{21}) (1 - \pi^*_1) \pi^*_1} \\
     &\xi = \frac{(\hat{\pi}^{*W}_{21} + \hat{\pi}^{*W}_{12})}{(1 - \hat{\pi}^{*W}_{12} - \hat{\pi}^{*W}_{21})},
\end{align}
where $\pi^*_1 = P(M^*_i = 1)$ is the empirical response probability of the observed mediator, $\hat{\pi}^{*W}_{21} = 1 - \hat{\pi}^{*W}_{11}$, and $\hat{\pi}^{*W}_{12} = 1 - \hat{\pi}^{*W}_{22}$.

The intercept of the \textit{outcome mechanism} is estimated as
\begin{align}
    \hat{\theta}_0 = \bar{Y} - \hat{\theta}_M \frac{\bar{M}^* - \hat{\pi}^{*W}_{12}}{(1 - \hat{\pi}^{*W}_{12} - \hat{\pi}^{*W}_{21})} - \bar{D}^T\hat{\theta}_D.
\end{align}

Jackknife or bootstrap procedures can be used to obtain standard errors for the estimated parameters.~\cite{efron1992bootstrap}

\subsection{Predictive Value Weighting}\label{pvw}
As in Section \ref{ols-correction}, the first step of the predictive value weighting (PVW) approach is to estimate the \textit{true mediator mechanism} and the \textit{observed mediator mechanisms} using the EM algorithm approach described in Webb and Wells (2023).\cite{hochstedler2023statistical} Again, we denote these estimates as $\hat{\boldsymbol{\beta}}^W$ and $\hat{\boldsymbol{\gamma}}^W$.

 We propose using the predictive value weighting (PVW) approach, introduced by Lyles and Lin (2010), to estimate the \textit{outcome model} in (\ref{eq:outcome-mechanism}) while accounting for the misclassified binary mediator $M$.\cite{lyles2010sensitivity} PVW is a method that adjusts for misclassified predictor variables in generalized linear regression models. The PVW approach uses an expanded dataset to fit a weighted regression model, where weights are computed using positive and negative predictive values. These predictive values are computed as a function of investigator-supplied, fixed sensitivity and specificity parameters. Our innovation is using the $\hat{\boldsymbol{\beta}}^W$ and $\hat{\boldsymbol{\gamma}}^W$ estimates from the EM algorithm analysis to estimate mediator sensitivity and specificity, rather than relying on fixed, investigator-supplied values as in Lyles and Lin (2010).\cite{lyles2010sensitivity} We compute the mediator sensitivity and specificity values for all individual subjects $i = 1, \dots, N$, denoted $\hat{\pi}^*_{i11}$ and $\hat{\pi}^*_{i22}$ respectively, in the dataset as follows:
\begin{align}
\label{eq:pistar-pvw-est}
\hat{\pi}^{*W}_{i11} = \frac{\text{exp}\{\hat{\gamma}^W_{110} + \boldsymbol{\hat{\gamma}^W_{11Z} Z_i}\}}{1 + \text{exp}\{\hat{\gamma}^W_{110} + \boldsymbol{\hat{\gamma}^W_{11Z} Z_i}\}},\\
\hat{\pi}^{*W}_{i22} = \frac{1}{1 + \text{exp}\{\hat{\gamma}^W_{120} + \boldsymbol{\hat{\gamma}^W_{12Z} Z_i}\}}.
\end{align}

Next, we specify a logistic regression model for the observed and potentially misclassified mediator $M^*$ as a function of other observed variables $Y$, $X$, and $\boldsymbol{C}$. Following Valeri and VanderWeele (2014), it is recommended that cubic splines are used to fit this model.\cite{valeri2014estimation} We use this model to compute $P(M^*_i = 1|  Y, X, \boldsymbol{C})$ for every observation in the sample.

Based on these sensitivity and specificity estimates and the logistic model for $M^* | Y, X, \boldsymbol{C}$, we compute subject-level positive predictive value (PPV) and negative predictive value (NPV) estimates as follows:
\begin{align}
    &PPV_i = \frac{\Bigl[ \frac{(\hat{\pi}^{*W}_{i22} - 1) \times (P(M^*_i = 1 | Y_i, X_i, \boldsymbol{C_i}) - 1)}{\hat{\pi}^{*W}_{i22} \times P(M^*_i = 1 | Y_i, X_i, \boldsymbol{C_i})} - 1 \Bigr]}{\Bigl[ \Bigl( \frac{(\hat{\pi}^{*W}_{i11} - 1) \times P(M^*_i = 1 | Y_i, X_i, \boldsymbol{C_i})}{\hat{\pi}^{*W}_{i11} \times (P(M^*_i = 1 | Y_i, X_i, \boldsymbol{C_i}) - 1)} \Bigr) \Bigl( \frac{(\hat{\pi}^{*W}_{i22} - 1) \times (P(M^*_i = 1 | Y_i, X_i, \boldsymbol{C_i}) - 1)}{\hat{\pi}^{*W}_{i22} \times P(M^*_i = 1 | Y_i, X_i, \boldsymbol{C_i})} \Bigr) - 1 \Bigr]} \\
    &NPV_i = \frac{\Bigl[ \frac{(\hat{\pi}^{*W}_{i11} - 1) \times P(M^*_i = 1 | Y_i, X_i, \boldsymbol{C_i})}{\hat{\pi}^{*W}_{i11} \times (P(M^*_i = 1 | Y_i, X_i, \boldsymbol{C_i}) - 1)} - 1 \Bigr]}{\Bigl[ \Bigl( \frac{(\hat{\pi}^{*W}_{i11} - 1) \times P(M^*_i = 1 | Y_i, X_i, \boldsymbol{C_i})}{\hat{\pi}^{*W}_{i11} \times (P(M^*_i = 1 | Y_i, X_i, \boldsymbol{C_i}) - 1)} \Bigr) \Bigl( \frac{(\hat{\pi}^{*W}_{i22} - 1) \times (P(M^*_i = 1 | Y_i, X_i, \boldsymbol{C_i}) - 1)}{\hat{\pi}^{*W}_{i22} \times P(M^*_i = 1 | Y_i, X_i, \boldsymbol{C_i})} \Bigr) - 1 \Bigr]}.
\end{align}

Next, we duplicate each record in the dataset and add a column for the unobserved true mediator variable, $M$. In the original record, we set $M = 0$. In the duplicate record, we set $M = 1$. We create a ``weight'' column $W$ in the new, combined dataset using the following specification:
\begin{align}
    &M_i = 1 \cap M^*_i = 1 \implies W_i = PPV_i \\
    &M_i = 0 \cap M^*_i = 1 \implies W_i = 1 - PPV_i \\
    &M_i = 1 \cap M^*_i = 0 \implies W_i = 1 - NPV_i \\
    &M_i = 0 \cap M^*_i = 0 \implies W_i = NPV_i.
\end{align}

Finally, to estimate the $\boldsymbol{\theta}$ parameters, we fit a weighted regression for $Y | M, X, \boldsymbol{C}$ using the duplicated dataset, with weights equal to $W_i$. Jackknife or bootstrap procedures can be used to compute standard errors for the $\boldsymbol{\beta}$, $\boldsymbol{\gamma}$, and $\boldsymbol{\theta}$ terms.

\subsection{Maximization Using an EM Algorithm}\label{em}
In this section, we describe the EM algorithm approach for estimating parameters $\boldsymbol{\beta}$, $\boldsymbol{\gamma}$, and $\boldsymbol{\theta}$ in (\ref{eq:true-mediator}) - (\ref{eq:outcome-mechanism}). We begin by constructing the complete data log-likelihood, where the true mediator $M$ is included as a latent variable:
\begin{equation}
    \begin{aligned}
    \label{eq:complete-log-like-em}
    \ell_{complete}(\boldsymbol{\beta},& \boldsymbol{\gamma}, \boldsymbol{\theta}; X, \boldsymbol{C}, \boldsymbol{Z}, Y) \\
    &= \sum_{i = 1}^N \bigg[ \ell_{Y | X, M, \boldsymbol{C}}(\boldsymbol{\theta}; X_i, M_i, \boldsymbol{C}_i, Y_i) +
    \ell_{M | X, \boldsymbol{C}}(\boldsymbol{\beta}; X_i, M_i, \boldsymbol{C}_i) +
    \ell_{M^* | M, \boldsymbol{Z}}(\boldsymbol{\gamma};M_i, \boldsymbol{Z}_i, M^*_i) \bigg] & \\
    &= \sum_{i = 1}^N \bigg[ \ell_{Y | X, M, \boldsymbol{C}}(\boldsymbol{\theta}; X_i, M_i, \boldsymbol{C}_i, Y_i) + \sum_{j = 1}^2 m_{ij} \text{log} \{ \pi_{ij} \} + \sum_{j = 1}^2 \sum_{\ell = 1}^2  m_{ij} m^*_{i\ell} \text{log} \{ \pi^*_{i \ell j} \}\bigg],
    \end{aligned}
\raisetag{12pt}\end{equation}
where $m_{ij} = \mathbbm{I}(M_i = j)$ and $m^*_{i\ell} = \mathbbm{I}(M^*_i = \ell)$. We write the contribution to this likelihood from the \textit{outcome mechanism} as a generic likelihood $\ell_{Y | X, M, \boldsymbol{C}}(\boldsymbol{\theta}; X_i, M_i, \boldsymbol{C}_i, Y_i)$ in order to preserve the generality of our algorithm. This choice introduces a slight abuse of notation. In (\ref{eq:complete-log-like-em}), the parameter vector $\boldsymbol{\theta}$ can now be considered a vector of all outcome regression parameters, including the variance $\sigma$, for example, if we model $Y$ as a Normal outcome.

For the expectation step of the algorithm, we express the expectation of $m_{ij}$ as
\begin{equation}
\begin{aligned}
\label{eq:e-step-em}
w_{ij} = P(M_i = j | M_i^*, X_i, \boldsymbol{C}_i, \boldsymbol{Z}_i, Y_i) = \sum_{\ell = 1}^2 \frac{m^*_{i\ell} \pi^*_{i\ell j} \pi_{ij} E[Y_i | X_i, M_i = j, \boldsymbol{C}_i, \boldsymbol{\theta}^{(t)}]}{\sum_{k = 1}^2 \pi^*_{i \ell k} \pi_{ik}E[Y_i | X_i, M_i = k, \boldsymbol{C}_i, \boldsymbol{\theta}^{(t)}]},
\end{aligned}
\end{equation}
where $t$ denotes the iteration of the EM algorithm.

In the maximization step, we maximize the following $Q$ function with respect to $\boldsymbol{\beta}$, $\boldsymbol{\gamma}$, and $\boldsymbol{\theta}$:
\begin{equation}
    \begin{aligned}
    \label{eq:q-em}
    Q
    = \sum_{i = 1}^N \big[ \sum_{j = 1}^2 \ell_{Y | X, M, \boldsymbol{C}}(\boldsymbol{\theta}; X_i, M_i = w_{ij}, \boldsymbol{C}_i, Y_i) + \sum_{j = 1}^2 w_{ij} \text{log} \{ \pi_{ij} \} + \sum_{j = 1}^2 \sum_{\ell = 1}^2  w_{ij} m^*_{i\ell} \text{log} \{ \pi^*_{i \ell j} \} \big].
    \end{aligned}
\end{equation}

Maximization of the $Q$ function in (\ref{eq:q-em}) can be conducted using four separate equations:
\begin{align}
\label{eq:q-beta}
 &Q_{\boldsymbol{\beta}} = \sum_{i = 1}^N \big[ \sum_{j = 1}^2 w_{ij} \text{log} \{ \pi_{ij} \} \big], \\
 \label{eq:q-gamma1}
 &Q_{\boldsymbol{\gamma}_{1}} = \sum_{i = 1}^N \big[ \sum_{\ell = 1}^2  w_{i1}  m^*_{i\ell} \text{log} \{ \pi^*_{i \ell 1} \} \big], \\
 \label{eq:q-gamma2}
 &Q_{\boldsymbol{\gamma}_{2}} = \sum_{i = 1}^N \big[ \sum_{\ell = 1}^2  w_{12}  m^*_{i\ell} \text{log} \{ \pi^*_{i \ell 2} \} \big], \\
 \label{eq:q-theta}
    &Q_{\boldsymbol{\theta}}
    = \sum_{i = 1}^N \big[ \sum_{j = 1}^2 \ell_{Y | X, M, \boldsymbol{C}}(\boldsymbol{\theta}; X_i, M_i = w_{ij}, \boldsymbol{C}_i, Y_i) \big].
\end{align}
$Q_{\boldsymbol{\beta}}$ denotes the component of the complete data log-likelihood corresponding to the \textit{true mediator mechanism} and containing the $\boldsymbol{\beta}$ parameters. $Q_{\boldsymbol{\gamma}_{1}}$ and $Q_{\boldsymbol{\gamma}_{2}}$ correspond to the sensitivity and specificity components of the \textit{observation mechanisms}, respectively. $Q_{\boldsymbol{\gamma}_{1}}$ contains all $\boldsymbol{\gamma}$ where the second subscript is $1$, corresponding to the case where $M = 1$. Likewise, $Q_{\boldsymbol{\gamma}_{2}}$ contains all $\boldsymbol{\gamma}$ where the second subscript is $2$, corresponding to the case where $M = 2$. $Q_{\boldsymbol{\theta}}$ contains the $\boldsymbol{\theta}$ parameters (possibly including a variance term, $\sigma$), and corresponds to the \textit{outcome mechanism} in (\ref{eq:outcome-mechanism}).

In practice, $Q_{\boldsymbol{\beta}}$ in (\ref{eq:q-beta}) can be fit as a standard logistic regression model where $w_{ij}$ (a value between 0 and 1) is input in place of the binary outcome. $Q_{\boldsymbol{\gamma}_{1}}$ and $Q_{\boldsymbol{\gamma}_{2}}$ in (\ref{eq:q-gamma1}) and (\ref{eq:q-gamma2}), respectively, are fit as weighted logistic regression models. The weights in these models are the E-step weights, $w_{i1}$ or $w_{i2}$, and the outcomes are the observed mediator indicator variable, $m^*_{i\ell}$.

The fitting procedure for $Q_{\boldsymbol{\theta}}$ in (\ref{eq:q-theta}) differs based on the assumed distribution of the outcome variable, $Y$. In the case of a Normal outcome variable, $Q_{\boldsymbol{\theta}}$ can be estimated using traditional maximum likelihood methods. For Binary and Poisson outcomes, $Q_{\boldsymbol{\theta}}$ can be estimated using a ``duplicated data'' approach. First, the dataset is duplicated and a column is added for the unobserved true mediator variable, $M$. We set $M = 0$ in the original record and $M = 1$ in the duplicate record. A column is also added for the ``weight'' term, denoted $W^{EM}$. When $M_i = 0$, this term is set to $w_{i2}$, indicating the reference category. When $M_i = 1$, $W_i^{EM} = w_{i1}$. To estimate $\boldsymbol{\theta}$, we fit a weighted generalized linear model to the duplicated dataset, using $Y$ as the outcome and $W^{EM}$ as the weights. We use the $M$ column in the duplicated data for the mediator variable in the model. Denote the parameter estimates resulting from this procedure as $\hat{\boldsymbol{\beta}}^{EM}$, $\hat{\boldsymbol{\gamma}}^{EM}$, and $\hat{\boldsymbol{\theta}}^{EM}$.

\subsubsection{Correcting for Label Switching}
In order to obtain the final parameters estimates for $\boldsymbol{\beta}$, $\boldsymbol{\gamma}$, and $\boldsymbol{\theta}$ using the EM algorithm, we must address the fundamental label switching problem inherent to this latent variable modeling approach. Label switching a known phenomenon where estimates of parameters $\boldsymbol{\alpha} = (\boldsymbol{\beta}, \boldsymbol{\gamma}_1, \boldsymbol{\gamma}_2, \theta_0, \theta_X, \theta_M, \boldsymbol{\theta}_{C}, \theta_{XM})$ are replaced with $\boldsymbol{\alpha}' = (-\boldsymbol{\beta}, \boldsymbol{\gamma}_2, \boldsymbol{\gamma}_1, \theta_0 + \theta_M, \theta_{X} + \theta_{XM}, -\theta_M, \boldsymbol{\theta}_C, -\theta_{XM})$ due to the permutation invariance of the complete data log-likelihood.\cite{collins2014estimation, hochstedler2023statistical} Thus, when we estimate the parameters using the proposed EM algorithm, it is possible that we estimated $\boldsymbol{\alpha}'$, the ``label-switched'' parameter set, rather than $\boldsymbol{\alpha}$, the parameter set of interest. Following Webb and Wells (2023), we use Youden's $J$ Statistic as a criteria to detect label switching.\cite{hochstedler2023statistical} The criteria and procedure for correcting label switching are described below.

First, we use $\hat{\boldsymbol{\gamma}}^{EM}$ to estimate average mediator sensitivity and specificity for all subjects in the dataset as follows: 
\begin{align}
\label{eq:avg_pistar-em}
\hat{\pi}^{*EM}_{11} = \frac{1}{N}\sum_{i = 1}^N \hat{\pi}^{*EM}_{i11} = \frac{1}{N}\sum_{i = 1}^N \frac{\text{exp}\{\hat{\gamma}^{EM}_{110} + \boldsymbol{\hat{\gamma}^{EM}_{11Z} Z_i}\}}{1 + \text{exp}\{\hat{\gamma}^{EM}_{110} + \boldsymbol{\hat{\gamma}^{EM}_{11Z} Z_i}\}},\\
\hat{\pi}^{*EM}_{22} = \frac{1}{N}\sum_{i = 1}^N \hat{\pi}^{*EM}_{i22} = \frac{1}{N}\sum_{i = 1}^N \frac{1}{1 + \text{exp}\{\hat{\gamma}^{EM}_{120} + \boldsymbol{\hat{\gamma}^{EM}_{12Z} Z_i}\}}.
\end{align}

Based on the relationship between $\boldsymbol{\alpha}$ and $\boldsymbol{\alpha}'$, the following relationship holds between the average mediator sensitivity and specificity values computed under the estimated and ``label switched'' parameter sets, denoted $\hat{\pi}^{*EM}_{jj}$ and $\hat{\pi}^{*switch}_{jj}$, respectively:
\begin{align}
\label{eq:pistar-switch}
    \hat{\pi}^{*switch}_{11} = 1 - \hat{\pi}^{*EM}_{22},
    \hat{\pi}^{*switch}_{22} = 1 - \hat{\pi}^{*EM}_{11}.
\end{align}
We use (\ref{eq:pistar-switch}) to estimate $\hat{\pi}^{*switch}_{11}$ and $\hat{\pi}^{*switch}_{22}$.

Next, we compute Youden's $J$ Statistic for both the EM and ``label switched'' $\hat{\pi}^{*}_{jj}$ estimates. Youden's $J$ Statistic is a measure of the performance of a classifier and takes values between $0$ and $1$. \cite{BERRAR2019546} For a perfect classifier, Youden's $J$ Statistic equals 1. Denote Youden's $J$ Statistic computed under $\hat{\pi}^{*EM}_{jj}$ and under $\hat{\pi}^{*switch}_{jj}$ as $\hat{J}^{EM}$ and $\hat{J}^{switch}$, respectively. $\hat{J}^{EM}$ and $\hat{J}^{switch}$ are computed as follows,
\begin{align}
    &\hat{J}^{EM} = \hat{\pi}^{*EM}_{11} + \hat{\pi}^{*EM}_{22} - 1 \\
    &\hat{J}^{switch} = \hat{\pi}^{*switch}_{11} + \hat{\pi}^{*switch}_{22} - 1.
\end{align}

If $\hat{J}^{EM} \geq \hat{J}^{switch}$, we assume that the EM algorithm estimated the parameter set of interest, $\boldsymbol{\alpha}$. If $\hat{J}^{EM} < \hat{J}^{switch}$, we assume that our estimated parameters actually correspond to the ``label-switched'' set, $\boldsymbol{\alpha}'$. Thus, if $\hat{J}^{EM} < \hat{J}^{switch}$ we have one additional step in our analysis to obtain the final parameter estimates for $\boldsymbol{\beta}$, $\boldsymbol{\gamma}$, and $\boldsymbol{\theta}$. We must apply the transformation to move between parameter set $\boldsymbol{\alpha}'$ and parameter set $\boldsymbol{\alpha}$. That is, we must multiply our $\hat{\boldsymbol{\beta}}^{EM}$, $\hat{\theta}_M^{EM}$, and $\hat{\theta}^{EM}_{XM}$ estimates by $-1$ and swap the indices between our $\hat{\boldsymbol{\gamma}}^{EM}_1$ and $\hat{\boldsymbol{\gamma}}^{EM}_2$ estimates. We must also add $\hat{\theta}^{EM}_M$ to the estimated intercept $\hat{\theta}^{EM}_0$ and add $\hat{\theta}^{EM}_{XM}$ to the estimated $\hat{\theta}^{EM}_X$ term. This proposed procedure relies on the assumption that the sum of the sensitivity and specificity of the mediator measurement instrument is greater than $1$, which is a common assumption used in other latent variable misclassification models to deal with label switching.\cite{collins2014estimation}

We invert the expected information matrix in order obtain the covariance matrix for $\boldsymbol{\beta}$, $\boldsymbol{\gamma}$, and $\boldsymbol{\theta}$. If label switching was present in the EM estimates, the rows and columns of the covariance matrix corresponding to $\hat{\boldsymbol{\gamma}}^{EM}_1$ and $\hat{\boldsymbol{\gamma}}^{EM}_2$ must also be relabeled by swapping the $1$ and $2$ indices for the $\boldsymbol{\gamma}$ terms. Covariance terms for $\boldsymbol{\beta}$ and $\hat{\theta}_M$ are unaffected by the sign change of the corresponding parameter estimates. 

\section{Simulations} \label{simulations}
We conducted simulation studies to evaluate the proposed ordinary least squares correction, predictive value weighting, and EM algorithm approaches in terms of bias and root mean squared error (rMSE). We compare these approaches to a naive mediation analysis that does not account for potential misclassification in the mediator variable. 

\begin{figure}[H]
\begin{center}
\includegraphics[scale=0.65]{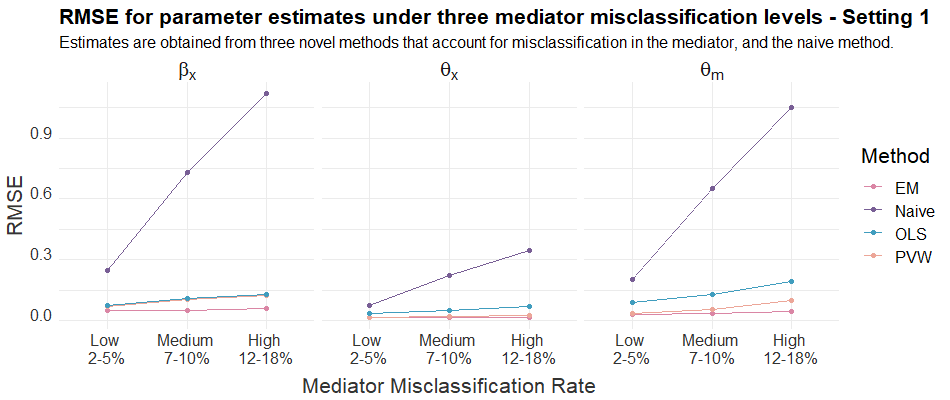}

\caption{Root mean squared error (RMSE) for three parameter estimates obtained from the EM algorithm, OLS correction, PVW, and naive analysis methods under three misclassification conditions in simulation setting 1.}\label{s1-figure}
\end{center}
\end{figure}

\begin{figure}[H]
\begin{center}
\includegraphics[scale=0.65]{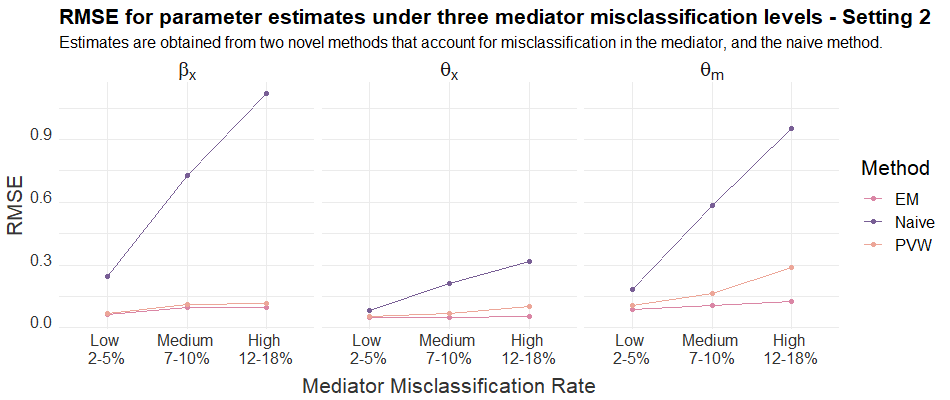}

\caption{Root mean squared error (RMSE) for three parameter estimates obtained from the EM algorithm, PVW, and naive analysis methods under three misclassification conditions in simulation setting 2.}\label{s2-figure}
\end{center}
\end{figure}

\begin{figure}[H]
\begin{center}
\includegraphics[scale=0.65]{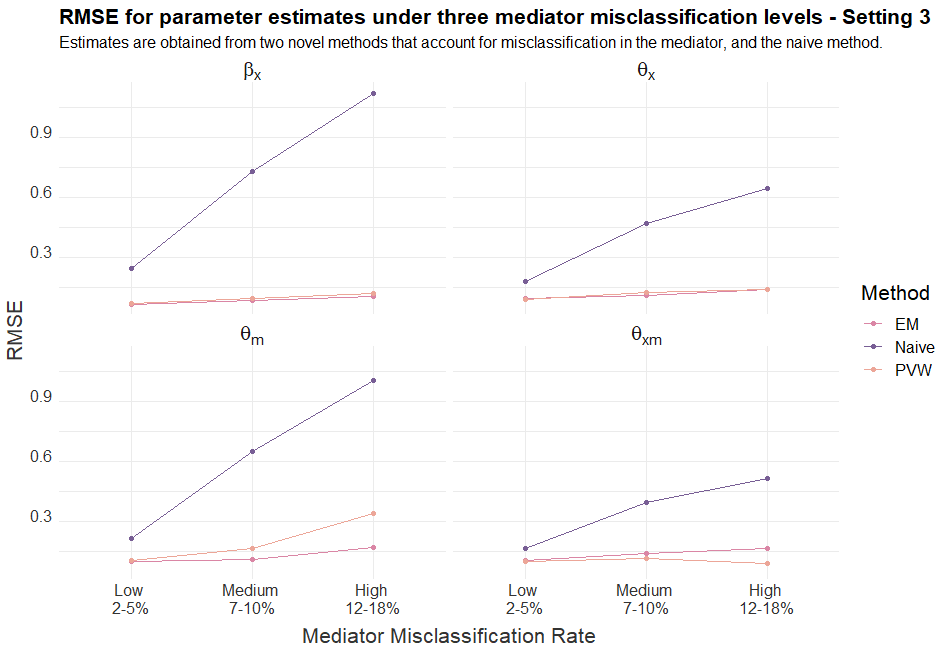}

\caption{Root mean squared error (RMSE) for four parameter estimates obtained from the EM algorithm, PVW, and naive analysis methods under three misclassification conditions in simulation setting 3.}\label{s3-figure}
\end{center}
\end{figure}

\begin{figure}[H]
\begin{center}
\includegraphics[scale=0.65]{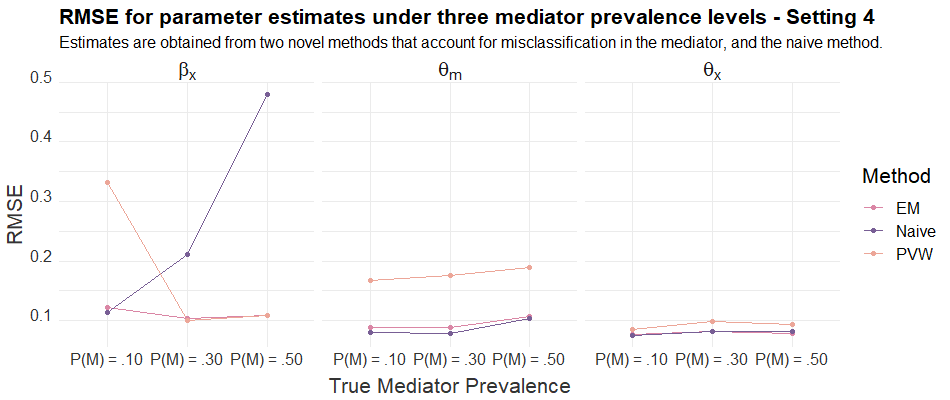}

\caption{Root mean squared error (RMSE) for four parameter estimates obtained from the EM algorithm, PVW, and naive analysis methods under three misclassification conditions in simulation setting 3.}\label{s4-figure}
\end{center}
\end{figure}

\begin{figure}[H]
\begin{center}
\includegraphics[scale=0.65]{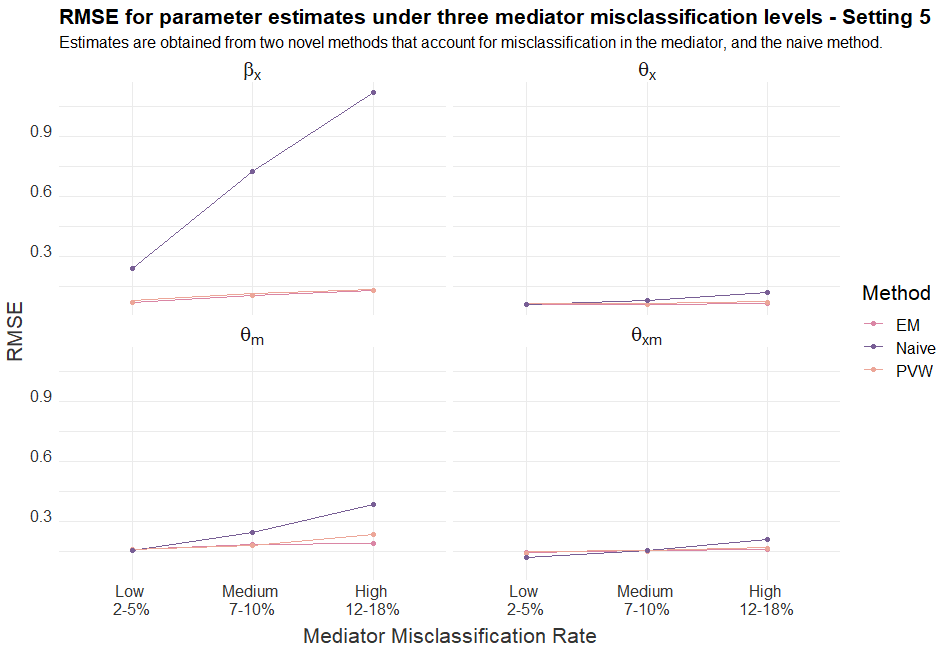}

\caption{Root mean squared error (RMSE) for four parameter estimates obtained from the EM algorithm, PVW, and naive analysis methods under three misclassification conditions in simulation setting 5.}\label{s5-figure}
\end{center}
\end{figure}

We investigated five simulation scenarios. First, we present results from a setting with a Normal outcome and no interaction term between $X$ and $M$. Next, we present results from a setting with a Bernoulli outcome and no interaction term between $X$ and $M$. The third simulation setting examines a case with a Bernoulli outcome and an interaction between $X$ and $M$. In our fourth simulation setting, we examine a setting with a Bernoulli outcome that is rare, perfect specificity in the measurement of $M$, and no interaction term between $X$ and $M$. Finally, we consider a simulation setting with a Poisson distributed outcome and an interaction term between $X$ and $M$. The bias and rMSE of selected parameter estimates from Setting 1 are presented in Table \ref{sim-results-table}. The bias and rMSE of selected parameter estimates from Settings 2 and 3 are presented in Table \ref{sim-results-table-2}. Results from Settings 4 and 5 are presented in Table \ref{sim-results-table-3}. Figures \ref{s1-figure}, \ref{s2-figure}, \ref{s3-figure}, \ref{s4-figure}, and \ref{s5-figure} graphically display the rMSE for selected parameters across differing misclassification rates in scenarios 1, 2, 3, 4, and 5, respectively. Details on the simulation settings are provided in Appendix \ref{appendix-sim-settings}. 

Across all simulation scenarios, the OLS correction, PVW, and EM algorithm approaches that account for misclassification outperformed the naive approach in terms of bias and rMSE. For the continuous outcome in setting 1, the OLS correction procedure performed slightly worse than PVW and the EM algorithm, likely due this method's use of average sensitivity and specificity estimates, rather than subject-level values. 

In settings 1, 2, 3, and 5, we explore the behavior of the proposed methods across different misclassification rates. In general, we observe that, as the mediator misclassification rate increases, bias and rMSE tend to increase for $\boldsymbol{\beta}$ and $\boldsymbol{\theta}$ parameter estimates. This is likely due to the added difficulty of estimating the true mediator model and outcome model in cases with more error. In contrast, the bias and rMSE of $\boldsymbol{\gamma}$ parameter estimates tends to decrease with increasing misclassification rates, likely due to the increasing information available for estimating the observed mediator model when more observations are subject to misclassification. 

In setting 4, we explore the behavior of the proposed PVW and EM algorithm approaches compared to the naive approach, across different prevalence levels of the binary mediator. The true mediator variable was either rare ($P(M = 1) \approx 0.10 - 0.15$), moderately common ($P(M = 1) \approx 0.25 - 0.30$), or common ($P(M = 1) \approx 0.50 - 0.55$). In this setting, the mediator was measured with perfect specificity and imperfect sensitivity ($P(M^* = 1 | M = 1) \approx 0.80 - 0.83$), but our methods allowed for the possibility of \textit{both} imperfect sensitivity and specificity. In addition, this setting had a relatively rare outcome variable ($P(Y = 1) \approx 0.10 - 0.20$). We also selected $\boldsymbol{Z}$ such that $\boldsymbol{Z} \subset \boldsymbol{c}$. This simulation scenario was designed because the low outcome prevalence, covariate subsetting, and perfect specificity would present a challenging, but realistic, estimation setting. These factors mirrored those in our applied study, and our goal was to test the performance of our proposed methods in a similar synthetic example. 

Across all mediator prevalence levels, the EM algorithm approach matched or outperformed the naive estimation method in terms of bias and rMSE. The naive method produced especially biased parameter estimates as the mediator prevalence, and thus as the magnitude of misclassification, increased. The PVW method generally produced estimates with larger bias and rMSE than the proposed EM algorithm. In addition, we found the PVW approach to be numerically unstable. Of the 500 simulated datasets, 1 produced such extreme estimates (i.e. $\beta_x > 500$) that it was removed from our numerical summary tables and plots. Even after removing this extreme realization, our results still show that the PVW approach generally performed worse than the EM algorithm and naive approaches in terms of rMSE. In setting 4, we generally observe that, as the true mediator prevalence increases, bias and rMSE tend to decrease for $\boldsymbol{\beta}$, $\boldsymbol{\gamma}$ and $\boldsymbol{\theta}$ parameter estimates. This trend is likely due to the increasing information available for estimating the observed mediator model, and particularly the sensitivity mechanism, when there are more observations such that $M = 1$.  

Across all settings, the EM algorithm approach generally outperformed the OLS correction and PVW procedures in terms of bias and rMSE. This finding is likely due to the seamless nature of the EM approach in comparison to the two-step OLS correction and PVW methods. In addition, it is notable that the EM algorithm was also more computationally efficient than the other two novel approaches. In simulation setting 1, for instance, the EM algorithm converged in approximately 6 seconds while the OLS correction and PVW approaches converged in 16 seconds and 18 seconds, respectively. For binary outcomes, the EM algorithm was slower to converge, but computation time was comparable to the PVW approach.

\section{Applied Example} \label{example} 
We applied the proposed predictive value weighting and EM algorithm approaches for misclassification of a binary mediator to a study of risk factors for pre-term birth using data from the National Vital Statistics System of the National Center for Health Statistics (NCHS).\cite{NCHS} Pre-term birth is an outcome of public health interest because infants born pre-term are known to experience more medical complications and developmental delays than those born full-term.\cite{mccormick1985contribution, kumar2008complications, hochstedler2021gestational} This study was inspired by the applied example in Valeri and VanderWeele (2014) investigating pre-eclampsia as a potential mediator of the relationship between maternal age and pre-term birth.\cite{valeri2014estimation}

We investigated gestational hypertension as a potential intermediate of the maternal age--pre-term birth relationship. Gestational hypertension, or high blood pressure during pregnancy, is estimated to affect between $10\%$ to $28\%$ of all pregnancies, depending on the diagnostic criteria used.\cite{bello2021prevalence} Maternal age has been found to be a risk factor for both pre-term delivery and for gestational hypertension.\cite{frick2021advanced} In addition, the risk of pre-term birth is higher among mothers with gestational hypertension, compared to mothers who do not have the condition.\cite{bertagnolli2016preterm, ofori2008risk}

We anticipate that gestational hypertension diagnosis may be subject to misclassification due to the wide variability in diagnostic criteria used to identify the condition.\cite{klemmensen2007validity} Previous studies have estimated sensitivity rates for pre-eclampsia, a related hypertensive condition, to be around $70\%$.\cite{klemmensen2005diagnostic} When considering all hypertensive disorders combined, sensitivity estimates drop below $50\%$.\cite{Callaghan2007invited} Specificity estimates for gestational hypertension are uniformly high ($99\%$ - $100\%$).\cite{Callaghan2007invited, klemmensen2005diagnostic} We anticipate that the imperfect sensitivity in previous studies may be tied to patient-level factors like race and insurance status. Insurance status may impact misdiagnosis rates because patients without insurance may not undergo costly diagnostic tests or see a medical professional as frequently as patients with health insurance.\cite{omalley2005measuring} In addition, previous studies have shown that Black mothers disproportionately report "near misses" in their medical care care, where health professionals fail to recognize serious complications during pregnancy and delivery.\cite{grant2023black} Among women with pregnancy-associated hypertension, Black mothers are more likely to be readmitted to the hospital due to their hypertension than mothers of other races. Taken together, these factors suggest that Black women's gestational hypertension symptom severity and diagnoses may not be fully captured in medical records.\cite{chornock2021racial} Recognizing this disparity, however, the American Heart Association recently stated stated that a personalized assessment of risk factors for gestational hypertension and related outcomes, in part based on patient race, is a ``rational approach''.\cite{garovic2022hypertension} There is hope that such recommendations lead to greater recognition of symptoms among Black mothers. 

We used our proposed methods to conduct a mediation analysis quantifying the effect of maternal age on pre-term birth through and independent of gestational hypertension using NCHS data. The original dataset included $3,669,928$ records derived from all birth certificates in the United States in the year 2021. We used a random subsample of $20,000$ observations to ensure reasonable computation time. Pre-term birth was defined as gestational age at birth below 37 weeks. Gestational hypertension was diagnosis criteria was unavailable.\cite{NCHS} Maternal age was centered and scaled before inclusion in the model. We adjusted for factors that may confound the association between maternal age and gestational hypertension, gestational hypertension and pre-term birth, and maternal and and pre-term birth. These factors included mother's race (self-reported as white or Black), whether the mother attended college, parity (a mother's number of previous live births), whether the mother used cigarette use during pregnancy, and mother's body mass index (BMI, centered and scaled). We modeled gestational hypertension diagnosis, conditional on true gestational hypertension, based on mother's race and insurance status (categorized as private or public vs. self-pay). We omitted records with missing or other values of any of the variables used in the analysis. We also limited the analysis to only include singleton births, cases with maternal age above $18$, and records with a parity of $8$ or less (records with parity above $8$ were removed because all observations with a parity above $8$ were coarsened into a single category in the original dataset).

In the sample of $20,000$ observations, the average maternal age was 29.1 years. A diagnosis of gestational hypertension was present in $9.5\%$ (1,908) of observations. $10.3\%$ of the deliveries in the dataset were before 37 weeks gestation, and thus categorized as pre-term. 

Table \ref{applied-results-table} presents parameter estimates for $\boldsymbol{\beta}$, $\boldsymbol{\gamma}$, and $\boldsymbol{\theta}$ terms from the proposed and naive estimation approaches. Using these results, we compute natural direct effect, natural indirect effect, and controlled direct effect estimates in Table \ref{applied-results-effect-table}.

In the naive approach, we run a logistic regression of gestational hypertension diagnosis on the predictor of interest and the remaining covariates. We estimate that maternal age slightly increases the risk of gestational hypertension ($\hat{\beta}_{age} = 0.077$, $\hat{OR}^{age} = 1.080$). We also ran a naive logistic regression of pre-term birth on maternal age, gestational hypertension diagnosis, and the remaining covariates. This analysis suggested that maternal age has a positive and significant total effect on pre-term birth ($\hat{OR}^{TE, age} = 1.108$). In addition, we found that gestational hypertension diagnosis was strongly associated with pre-term delivery ($\hat{\theta}_{GH^*} = 0.880$, $\hat{OR}^{GH^*} = 2.411$). The naive maternal age--pre-term birth interaction term failed to reach significance. Using these naive analysis results, we found that maternal age has a positive, significant direct effect on pre-term birth, independent of the gestational hypertension diagnosis pathway ($\hat{OR}^{NDE, age} = 1.121$). We also found a small positive and significant effect of maternal age on pre-term birth through gestational hypertension diagnosis ($\hat{OR}^{NIE, age} = 1.009$). 

The naive analysis approach does not account for misdiagnosis of gestational hypertension, which may yield bias in the parameter estimates and effect estimates. We conducted misclassification-corrected analyses using both the PVW and EM algorithm approaches. Both approaches suggest that the association between maternal age and true gestational hypertension is stronger than that estimated using the naive analysis (PVW: $\hat{\beta}_{age} = 0.091$, $\hat{OR}^{age} = 1.095$; EM: $\hat{\beta}_{age} = 0.103$, $\hat{OR}^{age} = 1.108$). In addition, the total effect estimate for the age--pre-term birth association is weaker in the analyses that account for misdiagnosis of gestational hypertension (PVW: $\hat{OR}^{TE, age} = 1.083$; EM: $\hat{OR}^{TE, age} = 1.054$), compared to appraoches that do not. The association between gestational hypertension and pre-term birth markedly strengthened using the PVW and EM algorithm approaches (PVW: $\hat{\theta}_{GH} = 1.243$, $\hat{OR}^{GH} = 3.466$; EM: $\hat{\theta}_{GH} = 1.188$, $\hat{OR}^{GH} = 3.281$), but the interaction term remained insignificant. With both misclassification-corrected procedures, we estimated that the natural direct effect and the natural indirect effect of maternal age on pre-term birth was larger than that of the naive analysis (PVW: $\hat{OR}^{NDE, age} = 1.131$, $\hat{OR}^{NIE, age} = 1.029$; EM: $\hat{OR}^{NDE, age} = 1.134$, $\hat{OR}^{NIE, age} = 1.034$). These results suggest that the association between maternal age and pre-term birth is not primarily driven by gestational hypertension status. 

The estimated $\boldsymbol{\gamma}$ parameters from the PVW and EM algorithm approaches allow us to compute sensitivity and specificity estimates for gestational hypertension diagnosis using (\ref{pistar-m}). Across all observations in the sample, the PVW $\hat{\boldsymbol{\gamma}}$ terms yield average sensitivity and specificity estimates of $37.70\%$ and $99.98\%$, respectively. Using the EM algorithm parameter estimates, we also estimate near perfect specificity and low sensitivity, at $99.86\%$ and $42.26\%$, respectively. The relatively large $\hat{\gamma}_{11,insurance}$ estimates from both misclassification-corrected approaches suggest that insurance status is related to the probability of gestational hypertension diagnosis, conditional on the condition being present (PVW: $\hat{\gamma}_{11,insurance} = -0.980$, $\hat{OR}^{insurance} = 0.375$; EM: $\hat{\gamma}_{11,insurance} = -1.006$, $\hat{OR}^{insurance} = 0.366$). Indeed, we find that estimated sensitivity differs greatly among patients who are insured vs. uninsured. Among insured individuals, we estimate a sensitivity of $38.45\%$ using PVW and $43.09\%$ using the EM algorithm. Among uninsured individuals, sensitivity estimates from the PVW and EM algorithm procedures are $18.96\%$ and $21.71\%$, respectively.

Using the proposed misclassification-corrected approaches, we found extreme parameter estimates and standard errors for the $\boldsymbol{\gamma}_{12}$ terms, which govern the specificity mechanism of the \textit{observed mediator model}. These parameter estimates yielded near-perfect specificity estimates in the sample. This perfect specificity result suggests that, by allowing for misclassification among individuals who truly do not have gestational hypertension, we have an incorrectly specified model. To test this theory, we estimated a model that did not include the specificity mechanism of the \textit{observed mediator model} using an adapted EM algorithm. Parameter and effect estimates from this method, denoted ``EM-Specificity'', can be found in Table \ref{applied-results-table} and Table \ref{applied-results-effect-table}, respectively. The results from the ``EM-Specificity'' method are nearly identical to those from other misclassification-corrected approaches. This finding suggests that the extreme $\boldsymbol{\gamma}_{12}$ estimates and standard errors obtained from the PVW and EM algorithm approaches are not detrimentally impacting estimation of other parameters in the model. Moreover, by computing sensitivity and specificity estimates using the estimated $\boldsymbol{\gamma}$ terms from a given method, an analyst can detect if a perfect sensitivity or specificity assumption may be valid, and account for this assumption as needed.

\section{Discussion} \label{discussion}
Misclassified binary mediator variables can severely bias parameter estimates and resulting causal effect estimates in a mediation analysis. In this paper, we developed three approaches to account for a misclassified binary mediator variable when gold standard labels are not available. These methods included an ordinary least squares correction approach for Normal outcome models, a predictive value weighting procedure, and a seamless EM algorithm approach. The key innovation of these strategies is the use of a covariate related to the \textit{observed mediator mechanism} to estimate mediator misclassification rates. In addition, we made these methods available through an open-source R Package, \textit{COMMA}.~\cite{COMMA} We compared the performance of these methods at various misclassification rates through simulation studies. In addition, we applied the PVW and EM algorithm procedures to a study of the mediating role of gestational hypertension on the association between maternal age and pre-term delivery.

Our methods are attractive because they are easy to implement and do not require external information, validation samples, or repeated measures. Our findings also suggest that our methods can handle cases of perfect sensitivity or specificity without having to make assumptions about misclassification rates. 

While the proposed methods account for misclassification in binary mediator variables, we do not consider the possibility of measurement error in the exposure or outcome variables. Future work could combine other misclassification-correction approaches to handle additional error-prone variables in the system. Moreover, we focus exclusively on binary mediator variables. However, to our knowledge, the impact of misclassification in general categorical mediator variables has yet to be studied.

\subsection*{Acknowledgements}
Funding support for KHW was provided by the LinkedIn and Cornell Ann S. Bowers College of Computing and Information Science strategic partnership PhD Award. MW was supported by NIH awards U19AI111143-07 and 1P01-AI159402.

\subsection*{Data Accessibility}
The data used in the applied example are freely available for download from the Center for Disease Control and Prevention's Vital Statistics Online Data Portal. This online portal can be accessed at the following link: https://data.nber.org/nvss/natality/csv/2021/

\newpage

\bibliographystyle{ama_with_first_initial_only}
\bibliography{references}

\begin{thebibliography}{10}

\bibitem{baron1986moderator}
Baron R.~M., Kenny D.~A.. The moderator--mediator variable distinction in social psychological research: Conceptual, strategic, and statistical considerations.  {\it Journal of Personality and Social Psychology. } 1986;51:1173.

\bibitem{jung2021introduction}
Jung S.~J.. Introduction to mediation analysis and examples of its application to real-world data  {\it Journal of Preventive Medicine and Public Health. } 2021;54:166.

\bibitem{imai2011commentary}
Imai K., Jo~B., Stuart E.~A.. Commentary: Using potential outcomes to understand causal mediation analysis  {\it Multivariate Behavioral Research. } 2011;46:861--873.

\bibitem{li2020direct}
Li~Y., VanderWeele T.~J.. Direct effects under differential misclassification in outcomes, exposures, and mediators  {\it Journal of Causal Inference. } 2020;8:286--299.

\bibitem{jiang2015causal}
Jiang Z., VanderWeele T.~J.. Causal mediation analysis in the presence of a mismeasured outcome  {\it Epidemiology. } 2015;26:e8--e9.

\bibitem{valeri2017misclassified}
Valeri L., Reese S.~L., Zhao S., et al. Misclassified exposure in epigenetic mediation analyses. Does DNA methylation mediate effects of smoking on birthweight?  {\it Epigenomics. } 2017;9:253--265.

\bibitem{ogburn2012analytic}
Ogburn E.~L., VanderWeele T.~J.. Analytic results on the bias due to nondifferential misclassification of a binary mediator  {\it American Journal of Epidemiology. } 2012;176:555--561.

\bibitem{le2012quantification}
Cessie S., Debeij J., Rosendaal F.~R., Cannegieter S.~C., Vandenbroucke J.~P.. Quantification of bias in direct effects estimates due to different types of measurement error in the mediator  {\it Epidemiology. } 2012;23:551--560.

\bibitem{vanderweele2019simple}
VanderWeele T.~J., Li~Y.. Simple sensitivity analysis for differential measurement error  {\it American Journal of Epidemiology. } 2019;188:1823--1829.

\bibitem{li2007estimation}
Li~Y., Schneider J.~A., Bennett D.~A.. Estimation of the mediation effect with a binary mediator  {\it Statistics in Medicine. } 2007;26:3398--3414.

\bibitem{valeri2013mediation}
Valeri L., VanderWeele T.~J.. Mediation analysis allowing for exposure--mediator interactions and causal interpretation: Theoretical assumptions and implementation with SAS and SPSS macros.  {\it Psychological Methods. } 2013;18:137.

\bibitem{valeri2014estimation}
Valeri L., Vanderweele T.~J.. The estimation of direct and indirect causal effects in the presence of misclassified binary mediator  {\it Biostatistics. } 2014;15:498--512.

\bibitem{lotspeich2022efficient}
Lotspeich S.~C., Shepherd B.~E., Amorim G.~G., Shaw P.~A., Tao R.. Efficient odds ratio estimation under two-phase sampling using error-prone data from a multi-national HIV research cohort  {\it Biometrics. } 2022;78:1674--1685.

\bibitem{tang2015binary}
Tang L., Lyles R.~H., King C.~C., Celentano D.~D., Lo~Y.. Binary regression with differentially misclassified response and exposure variables  {\it Statistics in Medicine. } 2015;34:1605-1620.

\bibitem{tao2021efficient}
Tao R., Lotspeich S.~C., Amorim G., Shaw P.~A., Shepherd B.~E.. Efficient semiparametric inference for two-phase studies with outcome and covariate measurement errors  {\it Statistics in Medicine. } 2021;40:725--738.

\bibitem{nguimkeu2021regression}
Nguimkeu P., Rosenman R., Tennekoon V.. Regression with a misclassified binary regressor: Correcting for the hidden bias   2021.

\bibitem{lyles2010sensitivity}
Lyles R.~H., Lin J.. Sensitivity analysis for misclassification in logistic regression via likelihood methods and predictive value weighting  {\it Statistics in Medicine. } 2010;29:2297--2309.

\bibitem{magder1997logistic}
Magder L.~S., Hughes J.~P.. Logistic regression when the outcome is measured with uncertainty  {\it American Journal of Epidemiology. } 1997;146:195--203.

\bibitem{beesley2020statistical}
Beesley L.~J., Mukherjee B.. Statistical inference for association studies using electronic health records: handling both selection bias and outcome misclassification  {\it Biometrics. } 2022;78:214--226.

\bibitem{hochstedler2023statistical}
Webb K.~A.~H., Wells M.~T.. Statistical inference for association studies in the presence of binary outcome misclassification  {\it arXiv preprint arXiv:2303.10215. } 2023.

\bibitem{webb2023assessment}
Webb K.~A.~H., Riley S.~A., Wells M.~T.. An assessment of racial disparities in pretrial decision-making using misclassification models  {\it arXiv preprint arXiv:2309.08599. } 2023.

\bibitem{COMMA}
Webb K.~A.~H.. COMMA: Correcting Misclassified Mediation Analysis  \url{https://github.com/kimberlywebb/COMMA} 2024.
\newblock R package version 1.0.0.

\bibitem{efron1992bootstrap}
Efron B.. Bootstrap methods: another look at the jackknife  in {\it Breakthroughs in statistics: Methodology and distribution}:569--593Springer 1992.

\bibitem{collins2014estimation}
Collins J., Huynh M.. Estimation of diagnostic test accuracy without full verification: a review of latent class methods  {\it Statistics in Medicine. } 2014;33:4141--4169.

\bibitem{BERRAR2019546}
Berrar D.. Performance Measures for Binary Classification  in {\it Encyclopedia of Bioinformatics and Computational Biology} (Ranganathan S., Gribskov M., Nakai K., Schönbach C.. , eds.):546-560Oxford: Academic Press 2019.

\bibitem{NCHS}
NCHS . {\it National Center for Health Statistic}.
\newblock U.S. Department of Health and Human Services 2021.

\bibitem{mccormick1985contribution}
McCormick M.~C.. The contribution of low birth weight to infant mortality and childhood morbidity  {\it New England Journal of Medicine. } 1985;312:82--90.

\bibitem{kumar2008complications}
Kumar P., Suresh G.. Complications after preterm birth: An overview for emergency physicians  {\it Clinical Pediatric Emergency Medicine. } 2008;9:191--199.

\bibitem{hochstedler2021gestational}
Hochstedler K.~A., Bell G., Park H., et al. Gestational age at birth and risk of developmental delay: the upstate KIDS study  {\it American Journal of Perinatology. } 2021;38:1088--1095.

\bibitem{bello2021prevalence}
Bello N.~A., Zhou H., Cheetham T.~C., et al. Prevalence of hypertension among pregnant women when using the 2017 American College of Cardiology/American Heart Association blood pressure guidelines and association with maternal and fetal outcomes  {\it JAMA Network Open. } 2021;4:e213808--e213808.

\bibitem{frick2021advanced}
Frick A.~P.. Advanced maternal age and adverse pregnancy outcomes  {\it Best Practice \& Research Clinical Obstetrics \& Gynaecology. } 2021;70:92--100.

\bibitem{bertagnolli2016preterm}
Bertagnolli M., Luu T.~M., Lewandowski A.~J., Leeson P., Nuyt A.~M.. Preterm birth and hypertension: is there a link?  {\it Current Hypertension Reports. } 2016;18:1--8.

\bibitem{ofori2008risk}
Ofori B.~D., Le~Tiec M., B{\'e}rard A.. Risk factors associated with preterm birth according to gestational age at birth  {\it Pharmacoepidemiology and Drug Safety. } 2008;17:556--564.

\bibitem{klemmensen2007validity}
Klemmensen {\AA}.~K., Olsen S.~F., {\O}sterdal M.~L., Tabor A.. {Validity of Preeclampsia-related Diagnoses Recorded in a National Hospital Registry and in a Postpartum Interview of the Women}  {\it American Journal of Epidemiology. } 2007;166:117-124.

\bibitem{klemmensen2005diagnostic}
Klemmensen {\AA}.~K., Olsen S.~F., Wengel C.~M., Tabor A.. Diagnostic criteria and reporting procedures for pre-eclampsia: a national survey among obstetrical departments in Denmark  {\it European Journal of Obstetrics \& Gynecology and Reproductive Biology. } 2005;123:41--45.

\bibitem{Callaghan2007invited}
Callaghan W.~M.. {Invited Commentary: Identifying Women with Hypertension during Pregnancy—Is High Specificity Sufficient?}  {\it American Journal of Epidemiology. } 2007;166:125-127.

\bibitem{omalley2005measuring}
O'{M}alley K.~J., Cook K.~F., Price M.~D., Wildes K.~R., Hurdle J.~F., Ashton C.~M.. Measuring diagnoses: ICD code accuracy  {\it Health Services Research. } 2005;40:1620--1639.

\bibitem{grant2023black}
Grant P.~A.. {\it Black Women’s Lived Experiences with Maternal near Misses: A Qualitative Analysis}.
\newblock PhD thesisWilmington University (Delaware) 2023.

\bibitem{chornock2021racial}
Chornock R., Iqbal S.~N., Kawakita T.. Racial disparity in postpartum readmission due to hypertension among women with pregnancy-associated hypertension  {\it American Journal of Perinatology. } 2021;38:1297--1302.

\bibitem{garovic2022hypertension}
Garovic V.~D., Dechend R., Easterling T., et al. Hypertension in pregnancy: diagnosis, blood pressure goals, and pharmacotherapy: a scientific statement from the American Heart Association  {\it Hypertension. } 2022;79:e21--e41.

\end{thebibliography}

\newpage
\section{Tables}

\begin{table}[H]\footnotesize
\centering
\caption{Bias and root mean squared error (rMSE) for a selection of parameter estimates from 500 realizations of simulation Setting 1. Estimates marked with a ``-'' are not obtained by the given estimation method.} \label{sim-results-table}
\begin{threeparttable}
\begin{tabular}{llc rr rr rr rr rr rr}
\hline
\noalign{\vskip 0.2cm}
        & &       & \multicolumn{2}{c}{OLS Correction\tnote{1}}       &
        \multicolumn{2}{c}{PVW\tnote{2}}& 
        \multicolumn{2}{c}{EM\tnote{3}}& 
        \multicolumn{2}{c}{Naive Analysis\tnote{4}}\\
        \cline{4-11}
Scenario &       & Truth & \multicolumn{1}{c}{Bias} & \multicolumn{1}{c}{rMSE} & \multicolumn{1}{c}{Bias} & \multicolumn{1}{c}{rMSE} & \multicolumn{1}{c}{Bias} & \multicolumn{1}{c}{rMSE} & \multicolumn{1}{c}{Bias} & \multicolumn{1}{c}{rMSE} \\
\hline
(1) No int and $Y \sim N$    \\
Low misclassification & $\beta_X$ & -2 & -0.012 & 0.072 & -0.012 & 0.072 & -0.002 & 0.049 & 0.244 & 0.247 \\
      & $\gamma_{11Z}$ & 2 & 0.156 & 1.253 & 0.156 & 1.253 & 0.016 & 0.461 & - & -\\
      & $\gamma_{12Z}$ & -2.5 & -0.717 & 3.031 & -0.717 & 3.031 & -0.072 & 0.719 & - & -\\
      & $\theta_X$ & 1.5 & 0.018 & 0.035 & -0.002 & 0.017 & -0.002 & 0.014 & 0.074 & 0.075\\
      & $\theta_M$ & -2 & 0.048 & 0.089 & -0.004 & 0.038 & -0.002 & 0.032 & 0.201 & 0.203\\
      \\
Med. misclassification & $\beta_X$ & -2 & -0.010 & 0.105 & -0.010 & 0.105 & -0.001 & 0.050 & 0.729 & 0.730 \\
      & $\gamma_{11Z}$ & 1 & 0.023 & 0.191 & 0.023 & 0.191 & 0.009 & 0.116 & - & - \\
      & $\gamma_{12Z}$ & -1 & -0.120 & 0.546 & -0.120 & 0.546 & -0.010 & 0.195 & - & - \\
      & $\theta_X$ & 1.5 & -0.011 & 0.048 & 0.008 & 0.021 & -0.001 & 0.015 & 0.223 & 0.224\\
      & $\theta_M$ & -2 & -0.027 & 0.128 & 0.033 & 0.057 & -0.002 & 0.038 & 0.650 & 0.650\\
              \\
High misclassification & $\beta_X$ & -2 & -0.019 & 0.124 & -0.019 & 0.124 & 0.002 & 0.062 & 1.118 & 1.118 \\
      & $\gamma_{11Z}$ & 1 & -0.000 & 0.119 & -0.000 & 0.119 & 0.001 & 0.077 & - & - \\
      & $\gamma_{12Z}$ & -1.5 & -0.019 & 0.303 & -0.019 & 0.303 & -0.012 & 0.173 & - & - \\
      & $\theta_X$ & 1.5 & -0.032 & 0.071 & 0.020 & 0.028 & 0.001 & 0.017 & 0.345 & 0.345\\
      & $\theta_M$ & -2 & -0.100 & 0.195 & 0.088 & 0.098 & 0.002 & 0.048 & 1.048 & 1.048\\
\hline  
\end{tabular}
\begin{tablenotes}
\item[1] ``OLS Correction'' estimates were obtained using the \textit{COMBO} R Package and an ordinary least squares correction method described in Section \ref{ols-correction}. 
\item[2] ``PVW'' estimates were obtained using the \textit{COMBO} R Package and a predictive value weighting method described in Section \ref{pvw}. 
\item[3] ``EM'' estimates were obtained from the proposed expectation-maximization algorithm in Section \ref{em}.
\item[4] ``Naive Analysis'' estimates were obtained from a standard mediation analysis that assumes no misclassification in the mediator variable. 
\end{tablenotes}
\end{threeparttable}
\end{table}

\begin{table}[H]\footnotesize
\centering
\caption{Bias and root mean squared error (rMSE) for a selection of parameter estimates from 500 realizations of simulation Settings 2 and 3. Estimates marked with a ``-'' are not obtained by the given estimation method.} \label{sim-results-table-2}
\begin{threeparttable}
\begin{tabular}{llc rr rr rr rr rr}
\hline
\noalign{\vskip 0.2cm}
        & &  &
        \multicolumn{2}{c}{PVW\tnote{1}}& 
        \multicolumn{2}{c}{EM\tnote{2}}& 
        \multicolumn{2}{c}{Naive Analysis\tnote{3}}\\
        \cline{4-9}
Scenario &       & Truth & \multicolumn{1}{c}{Bias} & \multicolumn{1}{c}{rMSE} & \multicolumn{1}{c}{Bias} & \multicolumn{1}{c}{rMSE} & \multicolumn{1}{c}{Bias} & \multicolumn{1}{c}{rMSE} \\
\hline
(2) No int and $Y \sim$ Bern    \\
Low misclassification & $\beta_X$ & -2 & -0.008 & 0.068 & -0.004 & 0.064 & 0.241 & 0.244\\
      & $\gamma_{11Z}$ & 2 & 0.229 & 1.110 & 0.182 & 0.927 & - & -\\
      & $\gamma_{12Z}$ & -2.5   & -0.726 & 4.241 & -0.536 & 2.345 & - & - \\
      & $\theta_X$ & 1.5 & -0.020 & 0.108 & -0.003 & 0.048 & 0.071 & 0.082\\
      & $\theta_M$ & -2  & -0.050 & 0.067 & -0.008 & 0.088 & 0.172 & 0.185\\
      \\
Med. misclassification & $\beta_X$ & -2 & -0.014 & 0.110 & -0.009 & 0.097 & 0.726 & 0.727\\
      & $\gamma_{11Z}$ & 1 & -0.007 & 0.173 & -0.009 & 0.156 & - & -\\
      & $\gamma_{12Z}$ & -1  & -0.093 & 0.473 & -0.072 & 0.426 & - & - \\
      & $\theta_X$ & 1.5 & -0.038 & 0.068 & 0.002 & 0.052 & 0.209 & 0.213\\
      & $\theta_M$ & -2  & -0.111 & 0.167 & -0.005 & 0.109 & 0.581 & 0.584\\
      \\
High misclassification & $\beta_X$ & -2 &  -0.012 & 0.118 & -0.002 & 0.080 & 1.117 & 1.118\\
      & $\gamma_{11Z}$ & 1 &  0.017 & 0.123 & 0.012 & 0.110 & - & -\\
      & $\gamma_{12Z}$ & -1.5 & -0.046 & 0.306 & -0.041 & 0.283 & - & - \\
      & $\theta_X$ & 1.5 & -0.089 & 0.104 & -0.003 & 0.053 & 0.313 & 0.315\\
      & $\theta_M$ & -2 & -0.250 & 0.287 & -0.007 & 0.127 & 0.949 & 0.950\\
      \\
(3) Int, $Y \sim$ Bern    \\
Low misclass. & $\beta_X$ & -2 & -0.010 & 0.070 & -0.009 & 0.066 & 0.239 & 0.243 \\
      & $\gamma_{11Z}$ & 2 & 0.363 & 1.410 & 0.271 & 1.080 & - & - \\
      & $\gamma_{12Z}$ & -2.5 & -0.816 & 3.059 & -0.699 & 2.822 & - & - \\
      & $\theta_X$ & 1.5 & -0.011 & 0.096 & -0.004 & 0.101 & 0.161 & 0.183 \\
      & $\theta_M$ & -2 & -0.034 & 0.105 & -0.007 & 0.097 & 0.203 & 0.215\\
      & $\theta_{XM}$ & 0.5 & -0.015 & 0.107 & -0.002 & 0.070 & -0.134 & 0.170\\
      \\
Med. misclass. & $\beta_X$ & -2 & -0.017 & 0.103 & -0.017 & 0.092 & 0.729 & 0.730\\
      & $\gamma_{11Z}$ & 1 & 0.007 & 0.174 & 0.005 & 0.140 & - & - \\
      & $\gamma_{12Z}$ & -1 & -0.094 & 0.511 & -0.078 & 0.453 & - & - \\
       & $\theta_X$ & 1.5 & -0.065 & 0.118 & -0.007 & 0.118 & 0.469 & 0.476\\
      & $\theta_M$ & -2 & -0.114 & 0.173 & -0.013 & 0.126 & 0.650 & 0.653\\
      & $\theta_{XM}$ & 0.5 & 0.032 & 0.108 & 0.005 & 0.140 & -0.387 & 0.399\\
              \\
High misclass. & $\beta_X$ & -2 & -0.006 & 0.116 & -0.003 & 0.103 & 1.120 & 1.120\\
      & $\gamma_{11Z}$ & 1 & 0.014 & 0.127 & 0.013 & 0.118 & - & - \\
      & $\gamma_{12Z}$ & -1.5 & -0.040 & 0.324 & -0.033 & 0.302 & - & - \\
       & $\theta_X$ & 1.5 & -0.095 & 0.133 & -0.001 & 0.141 & 0.641 & 0.645\\
      & $\theta_M$ & -2 & -0.273 & 0.317 & -0.018 & 0.164 & 1.007 & 1.009\\
      & $\theta_{XM}$ & 0.5 & 0.013 & 0.096 & 0.002 & 0.166 & -0.500 & 0.508\\
\hline  
\end{tabular}
\begin{tablenotes}
\item[1] ``PVW'' estimates were obtained using the \textit{COMBO} R Package and a predictive value weighting method described in Section \ref{pvw}. 
\item[2] ``EM'' estimates were obtained from the proposed expectation-maximization algorithm in Section \ref{em}.
\item[3] ``Naive Analysis'' estimates were obtained from a standard mediation analysis that assumes no misclassification in the mediator variable. 
\end{tablenotes}
\end{threeparttable}
\end{table}

\begin{table}[H]\footnotesize
\centering
\caption{Bias and root mean squared error (rMSE) for a selection of parameter estimates from 500 realizations of simulation Setting 4 and Setting 5. Estimates marked with a ``-'' are not obtained by the given estimation method.} \label{sim-results-table-3}
\begin{threeparttable}
\begin{tabular}{llc rr rr rr rr rr}
\hline
\noalign{\vskip 0.2cm}
        & &  &
        \multicolumn{2}{c}{PVW\tnote{1}}& 
        \multicolumn{2}{c}{EM\tnote{2}}& 
        \multicolumn{2}{c}{Naive Analysis\tnote{3}}\\
        \cline{4-9}
Scenario &       & Truth & \multicolumn{1}{c}{Bias} & \multicolumn{1}{c}{rMSE} & \multicolumn{1}{c}{Bias} & \multicolumn{1}{c}{rMSE} & \multicolumn{1}{c}{Bias} & \multicolumn{1}{c}{rMSE} \\
\hline
(4) No int, $Y \sim$ Bern,\\
\hspace{1.2em} perfect specificity \\
$P(M) = 0.10$ & $\beta_X$ & 2 & 0.025 & 0.332 & 0.038 & 0.123 & -0.073 & 0.114 \\
      & $\gamma_{11Z_1}$ & 0.5 & 0.979 & 39.781 & -0.296 & 2.465 & - & - \\
      & $\gamma_{11Z_2}$ & 0.1 & -5.320 & 67.531 & -1.372 & 9.101 & - & - \\
      & $\theta_X$ & 1.5 & 0.034 & 0.084 & 0.004 & 0.078 & 0.017 & 0.075 \\
      & $\theta_M$ & 0.2 & -0.130 & 0.167 & -0.004 & 0.089 & -0.019 & 0.080\\
      \\
$P(M) = 0.30$ & $\beta_X$ & 2 & 0.041 & 0.100 & 0.049 & 0.104 & -0.205 & 0.212\\
      & $\gamma_{11Z_1}$ & 0.5 & -0.362 & 5.646 & 0.017 & 0.191 & - & - \\
       & $\gamma_{11Z_2}$ & 0.1 & 0.901 & 9.608 & 1.524 & 4.337 & - & - \\
       & $\theta_X$ & 1.5 & 0.054 & 0.098 & -0.002 & 0.082 & 0.027 & 0.082\\
      & $\theta_M$ & 0.2 & -0.151 & 0.176 & 0.004 & 0.088 & -0.044 & 0.079\\
              \\
$P(M) = 0.50$ & $\beta_X$ & 2 & 0.058 & 0.109 & 0.060 & 0.110 & -0.478 & 0.480\\
      & $\gamma_{11Z_1}$ & 0.5 & 0.009 & 0.072 & 0.010 & 0.072 & - & - \\
       & $\gamma_{11Z_2}$ & 0.1 & 0.305 & 1.907 & 0.299 & 1.886 & - & - \\
       & $\theta_X$ & 1.5 & 0.056 & 0.094 & -0.004 & 0.078 & 0.032 & 0.081\\
      & $\theta_M$ & 0.2 & -0.177 & 0.189 & 0.010 & 0.107 & -0.082 & 0.103\\
(5) Int, $Y \sim$ Poisson    \\
Low misclass. & $\beta_X$ & -2 & -0.009 & 0.073 & -0.009 & 0.072 & 0.241 & 0.244 \\
      & $\gamma_{11Z}$ & 2 & 0.174 & 1.001 & 0.174 & 0.998 & - & - \\
      & $\gamma_{12Z}$ & -2.5 & -0.694 & 3.343 & -0.620 & 2.495 & - & - \\
      & $\theta_X$ & 1 & -0.004 & 0.061 & -0.002 & 0.061 & 0.012 & 0.061 \\
      & $\theta_M$ & -1 & -0.012 & 0.163 & -0.012 & 0.161 & 0.050 & 0.158\\
      & $\theta_{XM}$ & 0.5 & 0.011 & 0.142 & 0.006 & 0.146 & -0.028 & 0.123\\
      \\
Med. misclass. & $\beta_X$ & -2 & -0.007 & 0.108 & -0.007 & 0.106 & 0.727 & 0.728\\
      & $\gamma_{11Z}$ & 1 & 0.032 & 0.190 & 0.031 & 0.189 & - & - \\
      & $\gamma_{12Z}$ & -1 & -0.116 & 0.513 & -0.115 & 0.505 & - & - \\
       & $\theta_X$ & 1 & 0.006 & 0.062 & 0.002 & 0.063 & 0.061 & 0.082\\
      & $\theta_M$ & -1 & -0.001 & 0.183 & -0.004 & 0.184 & 0.201 & 0.248\\
      & $\theta_{XM}$ & 0.5 & -0.039 & 0.149 & -0.004 & 0.157 & -0.115 & 0.154\\
              \\
High misclass. & $\beta_X$ & -2 & -0.016 & 0.132 & -0.015 & 0.130 & 1.118 & 1.118\\
      & $\gamma_{11Z}$ & 1 & 0.005 & 0.117 & 0.005 & 0.116 & - & - \\
      & $\gamma_{12Z}$ & -1.5 & -0.021 & 0.312 & -0.021 & 0.310 & - & - \\
       & $\theta_X$ & 1 & 0.007 & 0.072 & -0.001 & 0.068 & 0.108 & 0.123\\
      & $\theta_M$ & -1 & -0.035 & 0.237 & -0.018 & 0.190 & 0.359 & 0.384\\
      & $\theta_{XM}$ & 0.5 & -0.053 & 0.165 & 0.006 & 0.163 & -0.190 & 0.211\\
\hline  
\end{tabular}
\begin{tablenotes}
\item[1] ``PVW'' estimates were obtained using the \textit{COMBO} R Package and a predictive value weighting method described in Section \ref{pvw}. 
\item[2] ``EM'' estimates were obtained from the proposed expectation-maximization algorithm in Section \ref{em}.
\item[3] ``Naive Analysis'' estimates were obtained from a standard mediation analysis that assumes no misclassification in the mediator variable. 
\end{tablenotes}
\end{threeparttable}
\end{table}

\begin{table}[h]
\centering
\caption{Parameter estimates and standard errors from the applied example using the national vital statistics dataset. All methods are described in Table \ref{sim-results-table} and Table \ref{sim-results-table-2}. Estimates marked with a ``-'' are not obtained by the given estimation method.} \label{applied-results-table}
\begin{threeparttable}
\begin{tabular}{ll rr rr rr rr rr rr}
\hline
           && 
           \multicolumn{2}{c}{PVW}       &
        \multicolumn{2}{c}{EM}       &
        \multicolumn{2}{c}{EM-Specificity} & 
        \multicolumn{2}{c}{Naive Analysis}\\
        \cline{3-10}
 && \multicolumn{1}{c}{Est.} & \multicolumn{1}{c}{SE} & \multicolumn{1}{c}{Est.} & \multicolumn{1}{c}{SE} & \multicolumn{1}{c}{Est.} & \multicolumn{1}{c}{SE} & \multicolumn{1}{c}{Est.} & \multicolumn{1}{c}{SE}\\
\hline
    \\
      $\beta_0$ && -0.870 & 0.321 & -1.036 & 0.233 & -1.053 & 0.246 & -2.101 & 0.047  \\
    $\beta_{age}$\tnote{1} && 0.091 & 0.048 & 0.103 & 0.037 & 0.104 & 0.036 & 0.077 & 0.028  \\
    $\beta_{race}$ && 0.052 & 0.525 & 0.042 & 0.349 & 0.171 & 0.264 & 0.082 & 0.062 \\
    $\beta_{education}$ && -0.037 & 0.087 & -0.036 & 0.076 & -0.039 & 0.072 & -0.026 & 0.058  \\
    $\beta_{parity}$ && -0.164 & 0.047 & -0.166 & 0.030 & -0.162 & 0.024 & -0.134 & 0.019 \\
    $\beta_{smoking}$ && -0.007 & 0.170 & -0.006 & 0.145 & -0.009 & 0.131 & -0.025 & 0.111 \\
    $\beta_{BMI}$ && 0.640 & 0.135 & 0.607 & 0.075 & 0.600 & 0.050 & 0.464 & 0.021  \\
    \\
       $\gamma_{110}$ && -0.476 & 1.248 & -0.274 & 0.312 & -0.259 & 0.696 & - & -   \\
       $\gamma_{11,race}$\tnote{2} && 0.033 & 1.061 & -0.026 & 0.819 & -0.085 & 0.870 &- & -   \\
       $\gamma_{11,insurance}$\tnote{3} && -0.980 & 1.328 & -1.006 & 0.395 & -0.885 & 0.585 & - & -   \\
       $\gamma_{120}$ && -36.581 & 10.075 & -9.878 & 11.525 & - & - & - & -   \\
       $\gamma_{12,race}$\tnote{2} && 16.515 & 9.698 & 4.843 & 7.738 &- & - & - & -   \\
       $\gamma_{12,insurance}$\tnote{3} && 17.084 & 10.668 & 2.086 & 8.814 & - & -  & - & -  \\
       \\
       $\theta_0$ && -2.703 & 0.207 & -2.648 & 0.138 & -2.637 & 0.052 & -2.342 & 0.047 \\
    $\theta_{age}$\tnote{4} && 0.080 & 0.041 & 0.017 & 0.053 & 0.013 & 0.030 & 0.103 & 0.029  \\
    $\theta_{race}$ && 0.538 & 0.155 & 0.541 & 0.106 & 0.510 & 0.084 & 0.534 & 0.056 \\
    $\theta_{education}$&& -0.408 & 0.065 & -0.400 & 0.064 & -0.399 & 0.064 & -0.401 & 0.059 \\
    $\theta_{parity}$ && 0.046 & 0.026 & 0.049 & 0.022 & 0.048 & 0.019 & 0.021 & 0.017 \\
    $\theta_{smoking}$ && 0.350 & 0.105 & 0.334 & 0.099 & 0.334 & 0.101 & 0.331 & 0.093  \\
    $\theta_{BMI}$ && -0.082 & 0.069 & -0.068 & 0.050 & -0.064 & 0.039 & 0.031 & 0.023 \\
    $\theta_{GH}$\tnote{5} && 1.243 & 0.269 & 1.188 & 0.173 & 1.171 & 0.054 & 0.880 & 0.064 \\
    $\theta_{GH \times age}$\tnote{6} && 0.072 & 0.061 & 0.194 & 0.089 & 0.201 & 0.084 & 0.057 & 0.062 \\
    \\
\hline  
\end{tabular}
\begin{tablenotes}
\item[1] $\beta_{age}$ refers to the association between maternal age (the predictor of interest) and true gestational hypertension status.
\item[2] $\gamma_{1,j,race}$ refers to the association between maternal race and gestational hypertension diagnosis, conditional on true gestational hypertension status.
\item[3] $\gamma_{1,j,insurance}$ refers to the association between insurance status and gestational hypertension diagnosis, conditional on true gestational hypertension status.
\item[4] $\theta_{age}$ refers to the association between maternal age (the predictor of interest) and preterm delivery.
\item[5] $\theta_{GH}$ refers to the association between true gestational hypertension status (the mediator) and preterm delivery.
\item[6] $\theta_{GH \times age}$ refers to the interaction term for true gestational hypertension status (the mediator) and maternal age (the predictor of interest).
\end{tablenotes}
\end{threeparttable}
\end{table}

\begin{table}[h]
\centering
\caption{Effect estimates and $95\%$ confidence intervals from the applied example using the national vital statistics dataset. All methods are described in Table \ref{sim-results-table} and Table \ref{sim-results-table-2}. } \label{applied-results-effect-table}
\begin{threeparttable}
\begin{tabular}{l rr rr rr rr }
\hline
           &
           \multicolumn{2}{c}{PVW}       &
        \multicolumn{2}{c}{EM}       &
        \multicolumn{2}{c}{EM-Specificity} & 
        \multicolumn{2}{c}{Naive Analysis}\\
        \cline{2-9}
 & \multicolumn{1}{l}{Est.} & \multicolumn{1}{r}{95\% CI} & \multicolumn{1}{l}{Est.} & \multicolumn{1}{r}{95\% CI} & \multicolumn{1}{l}{Est.} & \multicolumn{1}{r}{95\% CI} & \multicolumn{1}{l}{Est.} & \multicolumn{1}{r}{95\% CI}\\
\hline
\\
    \textit{NIE}\tnote{1} & 1.029 & (1.006, 1.090) & 1.034 & (1.013, 1.076) & 1.034 & (1.010, 1.064) & 1.009 & (1.002, 1.016) \\
   \textit{NDE}\tnote{2} & 1.131 & (1.078, 1.205) & 1.134 & (1.065, 1.217) & 1.132 & (1.061, 1.210) & 1.121 & (1.063, 1.183)\\
    \textit{CDE}\tnote{3} & 1.083 & (1.034, 1.217) & 1.017 & (0.905, 1.116) & 1.013 & (0.905, 1.116) & 1.109 & (1.047, 1.175)\\
    \\
\hline  
\end{tabular}
\begin{tablenotes}
\item[1] \textit{NIE} refers to the natural direct effect, which estimates the expected increase in the odds of preterm delivery as maternal age increases 1 standard deviation from the sample mean value, while fixing gestational hypertension status as negative.
\item[2] \textit{NDE} refers to the natural direct effect, which estimates the expected increase in the odds of preterm delivery when maternal age is fixed at the sample mean and gestational hypertension status changes to the value it would have attained for each individual for a one standard deviation increase in age.
\item[3] \textit{CDE} refers to the controlled direct effect, which estimates the expected increase in the odds of preterm delivery as maternal age increases 1 standard deviation from the sample mean value, conditioning on gestational hypertension status.
\end{tablenotes}
\end{threeparttable}
\end{table}

\newpage
\section{Figures}

\newpage

\begin{appendices}
\section{Simulation Settings}\label{appendix-sim-settings}
We present simulations for evaluating the proposed model with binary mediator misclassification correction in terms of bias and root mean squared error (rMSE). For a given simulation scenario, we present parameter estimates (for a subset of the parameters) obtained from the OLS correction method, the PVW approach, and our EM algorithm. We compare these approaches to a naive model that does not account for misclassification in the binary mediator variable. 

The OLS correction and PVW approaches require an EM algorithm for the first estimation step. For this EM algorithm, we used the ``squarem'' acceleration scheme from the \texttt{turboEM} package in \texttt{R}~\cite{stats2021R, turboEM}. We used a tolerance of $1 \times 10^{-7}$ for the absolute difference between successive log-likelihood values to assess converge. The alternative stopping criteria was reached when the maximum number of iterations (1500) was reached. For the proposed seamless EM algorithm approach, we used the same acceleration and convergence settings. All analyses were conducted in \texttt{R}~\cite{stats2021R}. 

For simulation settings 1, 2, 3, and 5, we generated $500$ datasets, each with a sample size of $10,000$. Average response misclassification probabilities for the simulation settings are provided in Table \ref{paper3-sim-settings-table}. 

\begin{table}[H]
\centering
\caption{$P(M = 1)$, $P(M^* = 1)$, $P(M^* = 1 | M = 1)$ (sensitivity), and $P(M^* = 2 | M = 2)$ (specificity) for simulation Settings 1, 2, 3, and 5.} \label{paper3-sim-settings-table}
\begin{threeparttable}
\begin{tabular}{clrrrrrr}
\hline
     Misclassification Level  && Quantity && Setting \\
\hline
\\
 Low     && $P(M = 1)$ & & 0.280 - 0.300 \\
              && $P(M^* = 1)$ && 0.280 - 0.300 \\
              && $P(M^* = 1 | M = 1)$ && 0.975 - 0.985 \\
              && $P(M^* = 2 | M = 2)$ && 0.955 - 0.965 \\
              \\
Medium   && $P(M = 1)$ & & 0.280 - 0.300 \\
              && $P(M^* = 1)$ && 0.300 - 0.315 \\
              && $P(M^* = 1 | M = 1)$ && 0.920 - 0.930 \\
              && $P(M^* = 2 | M = 2)$ && 0.890 - 0.910 \\
              \\
 High    && $P(M = 1)$ & & 0.280 - 0.300 \\
              && $P(M^* = 1)$ && 0.335 - 0.355 \\
              && $P(M^* = 1 | M = 1)$ && 0.840 - 0.860 \\
              && $P(M^* = 2 | M = 2)$ && 0.810 - 0.835 \\
              \\
\hline  
\end{tabular}
\end{threeparttable}
\end{table}

\subsection{Data Generation}
\subsubsection{Simulation Settings 1, 2, 3, and 5}
For each of the simulated datasets in settings 1, 2, 3, and 5, we first generated the predictor $X$ from a standard Normal distribution. $Z$ and $C$ were generated from Gamma distributions with shape parameters set to $1$. We used the following relationship to generate the true mediator value: $\text{logit}\{P(M = 1 | X, C)\} = 1 + (-2)X + (-2.5)C$.

The following relationships were used to generate the observed mediator for low, medium, and high misclassification rate settings for simulation scenarios 1, 2, and 3: 
\begin{align}\label{eq:paper3-appendix-mstar-gen}
\text{Low misclassification: } &\; \text{logit}\{ P(M^* = 1 | M = 1, Z) \} = 3 + (2)Z, \\
                                &\; \text{logit}\{ P(M^* = 1 | M = 2, Z) \} = (-2) + (-2.50)Z, \\
\text{Med. misclassification: } &\; \text{logit}\{ P(M^* = 1 | M = 1, Z) \} = 1.80 + (1)Z, \\
                                &\; \text{logit}\{ P(M^* = 1 | M = 2, Z) \} = (-1.50) + (-1)Z,\\
\text{High misclassification: } &\; \text{logit}\{ P(M^* = 1 | M = 1, Z) \} = 1 + (1)Z, \\
                                &\; \text{logit}\{ P(M^* = 1 | M = 2, Z) \} = (-0.50) + (-1.50)Z.
\end{align}

The following relationships were used to generate the outcome $Y$ for simulation settings 1, 2, 3, and 5, respectively:
\begin{align}\label{eq:paper3-appendix-y-gen}
\text{Setting 1: } &\; E(Y| X, M, C) = 1 + (1.50)X + (-2)M + (-0.20)C, \\
\text{Setting 2: } &\; \text{logit}\{ P(Y = 1 | X, M, C) \} = 1 + (1.50)X + (-2)M + (-0.20)C, \\
\text{Setting 3: } &\; \text{logit}\{ P(Y = 1| X, M, C) \} = 1 + (1.50)X + (-2)M + (-0.20)C + (0.50)XM, \\
\text{Setting 5: } &\; \text{log}\{Y | X, M, C) \} = -3 + (1)X + (-1)M + (-0.20)C + (0.50)XM.
\end{align}

\subsubsection{Simulation Setting 4}
Simulation setting 4 was designed to mirror a challenging and realistic scenario in applied settings. We generated data with a rare outcome, perfect specificity in the mediator, and with $\boldsymbol{Z} \subset \boldsymbol{C}$. In addition, we varied the prevalence of the true mediator, from approximately $0.10$ to $0.50$.

Predictor and confounding variables were generated as follows: 
\begin{align}\label{eq:paper3-appendix-s4-iv-gen}
    X \sim Bernoulli(0.67),\\
    C_1 \sim Gamma(1, 1),\\
    C_2 \sim \text{abs}(Normal(1,2)),\\
    C_3 \sim Bernoulli(0.20),\\
    C_4 \sim Bernoulli(0.55),\\
    C_5 \sim Normal(0,1),
\end{align}
where $\text{abs}(c)$ refers to the absolute value of $c$. In this setting, $Z_1 = C_1$ and $Z_2 = C_3$.

The following relationships were used to generate the true mediator for low, medium, and high prevalence level settings for simulation scenarios 4: $\text{logit}\{P(M = 1 | X, \textbf{C})\} = \beta_0 + (-2)X + (0.5)C_1 + (0)C_2 + (-2.50)C_3 +(-0.50)C_4 + (1)C_5$. The value of $\beta_0$ determined the mediator prevalence level. The values of $\beta_0$ were $-4, -2.5,$ and $-1$ for the low, medium, and high mediator prevalence levels, respectively.

The observed mediator was generated as follows: $\text{logit}\{ P(M^* = 1 | M = 1, \textbf{Z}) \} = 1 + (0.50)Z_1 + (0.10)Z_2$. To ensure perfect specificity, we generated data such that $P(M^* = 2 | M = 2, \textbf{Z}) = 1$ and $P(M^* = 1 | M = 2, \textbf{Z}) = 0$. 

Finally, we generated the outcome of interest as follows: $\text{logit}\{P(Y = 1 | X, \textbf{C}, M)\} = -4 + (1)C_1 + (0)C_2 + (0.50)C_3 + (-0.50)C_4 + (-2)C_5 + (1.50)X + (0.20)M$.

Simulation setting 4 had a sample size of 20,000. We generated 500 realizations for each mediator prevalence level.
\end{appendices}

\end{document}